\documentclass[a4paper,10pt]{revtex4}
\usepackage{mathrsfs}
\usepackage{graphicx}
\usepackage{latexsym}
\usepackage{amsmath}
\usepackage{amssymb}
\usepackage{textcomp}
\usepackage{amsbsy}
\usepackage{graphics}
\usepackage{epstopdf}
\usepackage{color}

\allowdisplaybreaks[4]

\begin{document}

\tolerance=5000

\title{Helical magnetogenesis with reheating phase from higher curvature coupling and baryogenesis}

\author{Kazuharu~Bamba,$^{1}$\,\thanks{bamba@sss.fukushima-u.ac.jp}
Sergei~D.~Odintsov,$^{2,3}$\,\thanks{odintsov@ieec.uab.es}
Tanmoy~Paul$^{4,5}$\,\thanks{pul.tnmy9@gmail.com}
Debaprasad~Maity$^{6}$\,\thanks{debu@iitg.ac.in}} \affiliation{ $^{1)}$ Division of Human Support System, Faculty of Symbiotic Systems Science, 
Fukushima University, Fukushima 960-1296, Japan \\ 
$^{2)}$ Institute of Space Sciences (IEEC-CSIC) C. Can Magrans s/n 08193 Bellaterra (Barcelona), Spain \\
$^{3)}$ ICREA, Passeig Luis Companys 23, 08010 Barcelona, Spain\\
$^{4)}$ Department of Physics, Chandernagore College, Hooghly - 712 136, India.\\
$^{5)}$ International Laboratory for Theoretical Cosmology, TUSUR, 634050 Tomsk, Russia.\\
$^{6)}$ Department of Physics, Indian Institute of Technology Guwahati, Assam 781039, India}

\tolerance=5000

\begin{abstract}
We investigate the generation of helical magnetic fields and address the baryon asymmetry of the universe from an 
inflationary magnetogenesis scenario, in which the conformal and parity symmetries of the electromagnetic field are 
broken through its coupling to the Ricci scalar and to the Gauss-Bonnet invariant via the dual field tensor, so that the generated 
magnetic field can have a helical nature. Depending on the reheating mechanism, we consider two different cases - (1) instantaneous reheating 
scenario, in which case the reheating e-fold number is zero, and (2) Kamionkowski reheating scenario which is parametrized by 
a non-zero e-fold number, a reheating equation of state parameter and a given reheating temperature. It 
is demonstrated that for both the reheating mechanisms, the generated magnetic fields can be compatible 
with the observations for suitable range of the model parameter present in the non-minimal coupling of the electromagnetic field. 
Actually the present magnetogenesis model does not produce sufficient hierarchy between the electric and 
magnetic fields at the end of inflation, and thus the electric field is not able to sufficiently induce (or enhance) the 
magnetic field during the Kamionkowski reheating stage. This in turn makes both the reheating cases almost similar from the perspective 
of magnetic field's evolution. Furthermore we find that the magnetic fields at the galactic scale with strength $\sim 10^{-13}\mathrm{G}$ 
can lead to the resultant value of the ratio of the baryonic number density to the entropy density as 
large as $\sim 10^{-10}$ , which is consistent with the observational data.
\end{abstract}


\maketitle

\section{Introduction}

Large scale magnetic field and baryon asymmetry of the universe are some of the important unresolved issues 
in the standard model of cosmology and particle physics. Despite of large attempts being made to understand those problems, 
it is still an active area of research. Although both the issues are seemingly disconnected, endeavour 
towards unified theoretical description would be extremely interesting. In order to proceed, one needs to go beyond standard model of 
cosmology and particle physics. Various astrophysical observations indicate that the universe is magnetized over a wide range of length scales. 
The magnetic fields have been detected in galaxies, galaxy clusters and even in intergalactic voids 
\cite{Grasso:2000wj,Beck:2000dc,Widrow:2002ud,Kandus:2010nw,Durrer:2013pga,Subramanian:2015lua}. 
On the other hand, it is observationally known that the universe carries 
a net excess of baryons over antibaryons \cite{Planck:2018jri,Riotto:1999yt,Giovannini:1997gp,Giovannini:1997eg,
Dine:2003ax,Cline:2006ts,Garbrecht:2018mrp,Odintsov:2016hgc,Bamba:2007hf,Bamba:2006km}. 
Such baryon asymmetry of the universe can be realized from the Sakharov's three conditions \cite{Sakharov:1967dj}: 
(1) baryon number non-conservation, (2) Charge(C) and Charge parity(CP) violation, and (3) departure from thermal equilibrium. 

Our current understanding on the origin of large scale magnetic fields are broadly classified into: (1) the astrophysical 
origin associated with dynamo mechanism \cite{Kulsrud:2007an,Brandenburg:2004jv,Subramanian:2009fu} 
and (2) the primordial origin where the magnetic field is generated from the primordial quantum 
fluctuations during the inflationary epoch \cite{Jain:2012ga,Durrer:2010mq,Kanno:2009ei,Campanelli:2008kh,
Demozzi:2009fu,Demozzi:2012wh,Bamba:2008ja,Bamba:2006ga,Bamba:2020qdj,Maity:2021qps,Haque:2020bip,
Giovannini:2021dso,Gasperini:1995dh,Kobayashi:2019uqs,Ratra:1991bn,Ade:2015cva,Chowdhury:2018mhj,Turner:1987bw,
Tripathy:2021sfb,Ferreira:2013sqa,Atmjeet:2014cxa,Kushwaha:2020nfa,Giovannini:2021thf,Adshead:2015pva}. 
The inflationary magnetogenesis earned significant attention due to the fact that the 
inflation  concomitantly solves the well known flatness and horizon problems and predicts almost scale invariant curvature perturbation which is in agreement with the Planck data 
\cite{guth,Linde:2005ht,Langlois:2004de,Riotto:2002yw,Baumann:2009ds} to a great precision. Due to this background curvature fluctuation the 
standard Maxwell's action  may produce magnetic field out of quantum vacuum, but its magnitude will naturally be very feeble and lie 
far below the observational constraints. The reason behind this feeble production of magnetic field is the conformal invariant nature of the standard 
electromagnetic theory and conformal flat nature of the background spacetime. Therefore, simplest way to enhance the strength of the magnetic field is to introduce the conformal non-invariant coupling in the electromagnetic sector. 
Several inflationary magnetogenesis models have been proposed so far, where the conformal non-invariant and non-minimal 
electromagnetic coupling has been introduced through the scalar field such as inflaton, axion and also through higher curvature terms
\cite{Jain:2012ga,Durrer:2010mq,Kanno:2009ei,Campanelli:2008kh,
	Demozzi:2009fu,Demozzi:2012wh,Bamba:2008ja,Bamba:2006ga,Bamba:2020qdj,Haque:2020bip,Giovannini:2021dso,
	Gasperini:1995dh,Kobayashi:2019uqs,Ratra:1991bn,Ade:2015cva,
	Chowdhury:2018mhj,Turner:1987bw,Tripathy:2021sfb,Ferreira:2013sqa,Atmjeet:2014cxa,Kushwaha:2020nfa,
	Adshead:2015pva,Caprini:2014mja,Kobayashi:2014sga,Atmjeet:2013yta,Fujita:2015iga,Campanelli:2015jfa,Tasinato:2014fia,Nandi:2021lpf}. 
Depending upon the basic physical requirements,those couplings can be classified  into: parity invariant and parity violating magnetogenesis scenario. 
Interestingly, parity violating and non-conformally coupled electromagnetic field  can generate helical magnetic field which has been shown to play 
crucial role in generating baryon asymmetry in the universe.

Apart from the inflationary magnetogenesis set-up, we should also mention a different proposal for magnetic field generation, 
in particular, from the bouncing scenario \cite{Frion:2020bxc,Chowdhury:2016aet,Koley:2016jdw,Qian:2016lbf}. 
Similar to inflation, bounce scenario is also able to generate sufficient strength of magnetic field 
at present universe. However bounce models are plagued with some problems, like the instability of curvature perturbation near the bounce, 
BKL instability, the dark energy issue etc \cite{Brandenberger:2012zb,Brandenberger:2016vhg,Battefeld:2014uga}. 
Here it may be mentioned that these severe problems can be solved to some extent in some modified 
theories of gravity \cite{Brandenberger:2012zb,Brandenberger:2016vhg,Battefeld:2014uga,Odintsov:2020zct,Odintsov:2021yva,Nojiri:2022xdo}. 

In regard to the relation between magnetogenesis and baryogenesis; it is well known that the presence of magnetic field leads to: 
(a) the dynamics of plasma to be out-of-thermal-equilibrium as in thermal equilibrium the photon distribution must be thermal 
which is incompatible with existence of long range magnetic field, (b) violation of time reversal (T) or in other words 
CP symmetry as well as C symmetry, and (c) being a vector quantity, the presence of magnetic field breaks the rotational invariance or SO(3) symmetry 
\cite{Davidson:1996rw}. Those are precisely the Sakharov's conditions for the successful generation of baryogenesis. 
Therefore, once we assume the genesis of magnetic 
field in the early universe that can act as a potential source for the production of baryon asymmetry in the universe. 
Motivated by this, Giovannini and Shaposhnikov \cite{Giovannini:1997gp,Giovannini:1997eg} first investigated the generation 
of baryon asymmetry through the abelian anomaly in the form of $U(1)$ Chern-Simons term which assumes non-zero vacuum expectation 
value due to helical magnetic field. This in turn produces net baryon asymmetry through the non-conservation of baryon number current. 
Related proposals are investigated in \cite{Joyce:1997uy,
	Thompson:1998qz,Vachaspati:1994ng,Rubakov:1996vz}. Hence, it is important to have non-zero helicity of the magnetic field 
	which is varying with the cosmological evolution. Axion- electrodynamics has been widely studied in this regard. Non-trivial 
	dynamics of axion field produces helical magnetic field which finally leads to the baryon asymmetry of the universe \cite{Turner:1987bw}. 
	Recently, the baryogenesis has been addressed from the production of helical magnetic fields, where 
the electromagnetic field gets coupled with the background Riemann tensor by the dual field tensor \cite{Kushwaha:2021csq}. Such coupling between the electromagnetic 
field and the Riemann tensor gives rise to helical nature of magnetic field, which in turn leads to a net baryonic number density as 
of the order $n_\mathrm{B}/s_0 \sim 10^{-10}$, that is consistent with the  cosmic microwave background (CMB) 
observations (where $n_\mathrm{B}$ is the net baryonic number density and 
$s_0$ is the entropy density of the present universe). From a different perspective, 
the article \cite{Giovannini:2021xbi} discussed the baryogenesis from magnetic fields, however from MHD amplification.

In the present work, we propose a helical magnetogenesis scenario from inflationary set-up and address the baryon asymmetry 
of the universe (BAU), from higher curvature coupling of the electromagnetic (EM) field. In particular, 
the EM field couples with the background higher curvature term(s), namely with the Ricci scalar and with the 
Gauss-Bonnet invariant, via the dual field tensor. The cosmology 
in Gauss-Bonnet gravity theory from various perspectives have been discussed in 
\cite{Li:2007jm,Carter:2005fu,Nojiri:2019dwl,Cognola:2006eg,Chakraborty:2018scm,Elizalde:2020zcb}. The presence of the 
non-minimal coupling in the EM action breaks the conformal and the 
parity symmetry, however preserves the U(1) invariance, of the EM field, which in turn leads to the gauge field production from the primordial 
Bunch-Davies vacuum. Moreover the positive and negative helicity modes 
of the EM field get different amplitudes along their cosmological evolution, and thus give rise to helical nature of the magnetic field. The 
Chern-Simons number stored in the helical magnetic fields is directly connected to the net baryonic number density of the universe. In regard to the 
the background spacetime, it is considered to be the $\alpha$-attractor scalar-tensor theory which is known 
to provide viable inflationary scenario consistent with the Planck results. 
Owing to the fact that the non-minimal coupling of the EM field 
occurs via the dual field tensor, the kinetic term of the EM field remains canonical even in presence of such non-minimal coupling, and thus the 
strong coupling problem is naturally resolved in the present magnetogenesis model. In such scenario, we explore the cosmological evolution 
of the EM field and consequently determine the net baryon density of the universe. 
Most of the earlier magnetogenesis scenarios did not explore the effect of reheating phase until recently.
Reheating can be important in this regard has recently been realized in \cite{Kobayashi:2019uqs}, 
and further explored in \cite{Bamba:2020qdj,Maity:2021qps,Haque:2020bip}. During the expansion, the universe enters into a reheating phase 
after inflation and depending on the reheating dynamics, we consider two different cases: (1) the instantaneous reheating case where 
the universe suddenly jumps to the radiation dominated epoch after the end of inflation, and thus the reheating era has zero e-fold number, and (2) the 
case where the reheating era possesses a non-zero e-fold number, in particular, we consider the Kamionkowski reheating mechanism that 
parametrizes the reheating dynamics by a non-zero e-fold number, a constant equation of state parameter and a given reheating temperature \cite{Dai:2014jja} 
(see also 
\cite{Cook:2015vqa,Albrecht:1982mp,Ellis:2015pla,Ueno:2016dim,Eshaghi:2016kne,Maity:2018qhi,Haque:2021dha,DiMarco:2017zek,Drewes:2017fmn,DiMarco:2018bnw}). 
These two cases make certain differences in the evolution of the helicity power spectrum. 
Here we would like to mention that the baryogenesis from helical magnetic fields with curvature couplings have been proposed earlier, 
however in quite different contexts \cite{Kushwaha:2021csq}. 
It may be noted that in our present analysis, we include the higher curvature Gauss-Bonnet coupling in the set-up 
and also discuss the possible effects of the reheating phase in the production of helical magnetic field and consequently in the baryogenesis, which makes
the present model essentially different from earlier ones.

The paper is organized as follows: in Sec.~\ref{sec_model}, we describe the model; in Sec.~\ref{sec_general expression}, 
we give the general expressions for the EM power spectra and the helicity spectrum in the 
present; Sec.~\ref{sec_solution_inflation} is reserved for the solution of the vector potential, the electromagnetic energy density 
and the helicity spectrum during inflation. The corresponding calculations during the reheating phase are carried out in 
Sec.~\ref{sec_instantaneous reheating} and Sec.~\ref{sec_kamionkowski}, which  correspond to the cases of  instantaneous reheating and a 
Kamionkowski like reheating model, respectively. The paper ends with some conclusions.

\section{The model}\label{sec_model}
The action is,
\begin{eqnarray}
 S = S_{grav} + S_{em}^{(can)} + S_{CB}~~,
 \label{action0}
\end{eqnarray}
where $S_{grav}$, namely the gravitational action, determines the background evolution and a non-minimally coupled electromagnetic (EM) field propagates 
over the background spacetime, where $S_{em}^{(can)}$ and $S_{CB}$ represent the canonical kinetic term and the non-minimal coupling function 
of the EM field respectively. The EM field couples with the higher curvature of the background spacetime, which actually provides 
the non-minimal coupling term in the action.

We focus on a specific inflationary model that permits slow roll inflation. In particular, we 
consider the so-called $\alpha$-attractor model, which unifies a large number of inflationary potentials \cite{Drewes:2017fmn}. 
If $\Phi$ is the canonical scalar field driving inflation, the model is described by the following action,
\begin{eqnarray}
 S_{grav} = \int d^4x\sqrt{-g}\bigg[\frac{R}{2\kappa^2} - \frac{1}{2}\partial_{\mu}\Phi\partial_{\nu}\Phi - V(\Phi)\bigg]~~,
 \label{action part 1}
\end{eqnarray}
where $\kappa^2 = 8\pi G$ ($G$ is the Newton's constant) and $V(\Phi)$ denotes the scalar field potential having the form,
\begin{eqnarray}
 V(\Phi) = \Lambda^4\left[1-\exp{\left(-\kappa\Phi\sqrt{\frac{2}{3\alpha}}\right)}\right]^{2n}~.
 \label{attractor potential}
\end{eqnarray}
Here the scale $\Lambda$ can be determined using the constraints from the observations of the 
anisotropies in the CMB. It is worth pointing out here that, for $\alpha = 1$ and $n = 1$, the above potential reduces to the well known 
Starobinsky model. We should also mention that the potential in Eq.(\ref{attractor potential}) contains a plateau at suitably 
large values of the field, which is favored by the CMB data. Finally, the $\alpha$-attractor inflationary model, i.e the $S_{grav}$, 
is known to provide viable inflationary scenario, with the observable quantities like the spectral index, tensor-to-scalar ratio 
being compatible with the Planck 2018 constraints \cite{Planck:2018jri}.

The canonical kinetic term of the EM field is given by,
\begin{eqnarray}
 S_{em}^{(can)} = \int d^4x\sqrt{-g}\big[-\frac{1}{4}F_{\mu\nu}F^{\mu\nu}\big]~~,
 \label{action part 2}
\end{eqnarray}
where $F_{\mu\nu} = \partial_{\mu}A_{\nu} - \partial_{\nu}A_{\mu}$ is the EM field tensor and $A_{\mu}$ is the corresponding field. 
In regard to the $S_{CB}$, as mentioned earlier, the EM field non-minimally 
couples with the spacetime curvature of the background spacetime, in particular,
\begin{eqnarray}
 S_{CB} = \int d^4x\sqrt{-g} f(R,\mathcal{G})\left[-\lambda F_{\mu\nu}\widetilde{F}^{\mu\nu}\right]~~,
 \label{actuon part 3}
\end{eqnarray}
where $\widetilde{F}^{\mu\nu} = \epsilon^{\mu\nu\alpha\beta}F_{\alpha\beta}$ with 
$\epsilon^{\mu\nu\alpha\beta}$ is the four dimensional Levi-Civita tensor and defined by $\epsilon^{\mu\nu\alpha\beta} 
= -\frac{1}{\sqrt{-g}}[\mu\nu\alpha\beta]$, the $[\mu\nu\alpha\beta]$ denotes 
the completely antisymmetric permutation having $[0123] = 1$. The EM coupling function 
$f(R,\mathcal{G})$ is considered to depend on the background Ricci scalar and Gauss-Bonnet curvature respectively, particularly the form of 
$f(R,\mathcal{G})$ is given by,
\begin{eqnarray}
 f(R,\mathcal{G}) = \kappa^{2q}\big(R^q + \mathcal{G}^{q/2}\big)~~,
 \label{form of f 1}
\end{eqnarray}
with $q$ being a parameter of the model. Later we will show that $q \sim \mathcal{O}(1)$ is compatible with the large scale observations of 
present magnetic strength. It may be observed that the EM field gets the non-minimal coupling via the dual field tensor, in particular by 
$F_{\mu\nu}\widetilde{F}^{\mu\nu}$ controlled by the dimensionless quantity $\lambda f(R,\mathcal{G})$, which 
acts as a parity violating agent in the EM field action and ensures the helical nature 
of the magnetic field. Here it deserves mentioning that the $S_{CB}$ spoils the conformal invariance of the EM field, although 
preserves the U(1) invariance of the gauge field. The broken conformal symmetry is the key ingredient in the magnetogenesis scenario, otherwise 
the EM field energy density redshifts as $1/a^4$ with the cosmological expansion of the universe and results to a very feeble magnetic strength at 
present epoch -- not compatible with the CMB observation at all. During the early universe when the spacetime curvature is high, the conformal breaking 
coupling leads to a non-trivial contribution to the EM field equations, however at late times (in particular at the end of inflation) 
the $S_{CB}$ will not contribute and consequently the EM field behaves as standard Maxwellian theory. Due to the presence of 
$F_{\mu\nu}\widetilde{F}^{\mu\nu} \left(= \epsilon^{\mu\nu\alpha\beta}F_{\mu\nu}F_{\alpha\beta}\right)$ 
in Eq.(\ref{actuon part 3}), the conformal breaking coupling 
does not contain the term like $(A_i')^2$ (where the prime denotes 
the derivative with respect to the conformal time) and thus the kinetic term of the EM field, i.e $(A_i')^2$, comes only through 
the $S_{em}^{(can)}$ given in Eq.(\ref{action part 2}). This in turn indicates that the EM kinetic term remains canonical in the present magnetogenesis 
model where the EM field gets coupled with the spacetime curvature through the corresponding dual field tensor. As a consequence, the model is free 
from the strong coupling problem. Moreover we will show later that the EM field energy density has negligible backreaction on the 
background inflationary FRW spacetime, which ensures the resolution of the backreaction issue in the present context.

In such magnetogenesis set-up, we aim to determine the magnetic strength at the present epoch and consequently the baryon asymmetry of the universe 
from the helical nature of the magnetic field. Varying the action (\ref{action0}) with respect to $A_{\mu}$, we get the field equation for the 
EM field as,
\begin{eqnarray}
 \partial_{\alpha}\left[\sqrt{-g}\left\{g^{\mu\alpha}g^{\nu\beta}F_{\mu\nu} 
 + 8\lambda f(R,\mathcal{G})~\epsilon^{\mu\nu\alpha\beta}\partial_{\mu}A_{\nu}\right\}\right] = 0~~,
 \label{eom}
\end{eqnarray}
where it may be noticed that the conformal breaking coupling modifies the EM field equation compared to the standard Maxwell's equation. A spatially 
flat FRW metric ansatz will fulfill our purpose in the present context, i.e
\begin{eqnarray}
 ds^2 = a^2(\eta)\big[-d\eta^2 + d\vec{x}^2\big]~~,
 \label{FRW metric ansatz}
\end{eqnarray}
where $a(\eta)$ is the scale factor of the universe with $\eta$ being the conformal time and related to the cosmic time ($t$) by 
$d\eta = \frac{dt}{a(\eta)}$. For this metric 
ansatz, the temporal component of the EM field Eq.(\ref{eom}) becomes,
\begin{eqnarray}
 \partial^{i}\big[\partial_iA_0 - \partial_0A_i\big] = 0~~.
 \label{temporal eom}
 \end{eqnarray}
 Due to the antisymmetric nature of $\epsilon^{\mu\nu\alpha\beta}$, the temporal component Eq.(\ref{temporal eom}) seems to free 
from the effect of the coupling function $f(R,\mathcal{G})$. Here we consider the Coulomb gauge condition, in particular $\partial_{i}A^{i} = 0$, 
due to which, Eq.(\ref{temporal eom}) leads to the solution as $A_0 = 0$. Therefore in effect of $\partial_{i}A^{i} = 0$ and $A_0 = 0$, 
the spatial component of the EM field Eq.(\ref{eom}) takes the following form,
\begin{eqnarray}
 A_i''(\eta,\vec{x}) - \partial_l\partial^{l}A_i + 8\lambda f'(R,\mathcal{G})~\epsilon_{ijk}\partial_{j}A_{k} = 0~~,
 \label{eom2}
\end{eqnarray}
where $\epsilon_{ijk} = \left[0ijk\right]$. As mentioned earlier, the $\alpha$-attractor model is able to drive successful inflationary 
scenario during the early universe for suitable values of $\alpha$ and $n$ respectively. 
Keeping this in mind, we consider a quasi-de-Sitter inflationary background spacetime 
in the present context, where the scale factor has the following form,
\begin{eqnarray}
 a(\eta) = \bigg(\frac{-\eta}{\eta_0}\bigg)^{\beta + 1}~~~~~~~~\mathrm{with}~~~~~~~~~\beta = -2-\epsilon = -3 + \frac{\mathcal{H}'}{\mathcal{H}^2}~~.
 \label{scale factor}
\end{eqnarray}
Here, and also in the subsequent calculation, a prime denotes $\frac{d}{d\eta}$, $\mathcal{H}$ is the conformal Hubble parameter and defined 
by $\mathcal{H} = a'/a$. Moreover $\epsilon$ is known as the slow roll parameter having the expression 
$\epsilon = -\frac{\mathcal{H}'}{\mathcal{H}^2} + 1$. Eq.(\ref{scale factor}) evidents that for $\epsilon = 0$, the scale factor 
becomes $a(\eta) \propto \left(-\eta\right)^{-1}$ which results to the de-Sitter evolution of the universe. The scale factor (\ref{scale factor}) 
immediately leads to the conformal Hubble parameter, Ricci scalar and the Gauss-Bonnet invariant as,
\begin{eqnarray}
 \mathcal{H} = \frac{\beta + 1}{\eta}
 \label{Hubble parameter}
\end{eqnarray}
and
\begin{eqnarray}
 R&=&\frac{6}{a^2}\big(\mathcal{H}' + \mathcal{H}^2\big) = \frac{6\beta(\beta + 1)}{\eta_0^2}\bigg(\frac{-\eta}{\eta_0}\bigg)^{2\epsilon}\nonumber\\
 \mathcal{G}&=&\frac{24}{a^4}\mathcal{H}^2\mathcal{H}' = -\frac{24(\beta + 1)^3}{\eta_0^4}\bigg(\frac{-\eta}{\eta_0}\bigg)^{4\epsilon},
 \label{R and G}
\end{eqnarray}
respectively. The cosmic Hubble parameter (defined by $H = \dot{a}/a$, where an overdot represents $\frac{d}{dt}$) is related to the conformal 
Hubble parameter as $H = \frac{1}{a}\mathcal{H}$ and thus we can express $H$ as
\begin{eqnarray}
 H = \frac{1}{\eta_0}\exp{\left(-\epsilon N\right)}~~,
 \label{cosmic Hubble parameter}
\end{eqnarray}
in terms of the e-folding number (symbolized by $N$). The e-folding number up-to the time $\eta$ is defined as $N = \int^{\eta} aH~d\eta$, 
where $N = 0$ refers the instance of the beginning of inflation which, in the present context, 
we consider to happen when the CMB scale mode ($\sim 0.05\mathrm{Mpc}^{-1}$) crosses the Hubble horizon. If $N_\mathrm{f}$ denotes the total 
e-fold number of inflation \cite{Drewes:2017fmn}, then
\begin{eqnarray}
 N_\mathrm{f} = \frac{3\alpha}{4n}\left[\exp{\left(\kappa\Phi_\mathrm{i}\sqrt{\frac{2}{3\alpha}}\right)} 
 - \exp{\left(\kappa\Phi_\mathrm{f}\sqrt{\frac{2}{3\alpha}}\right)} - \kappa\sqrt{\frac{2}{3\alpha}}\left(\Phi_\mathrm{i} - \Phi_\mathrm{f}\right)\right]~,
 \label{inflation-e-fold}
\end{eqnarray}
where $\Phi_\mathrm{i}$ and $\Phi_\mathrm{f}$ are given by,
\begin{eqnarray}
 \Phi_\mathrm{i}&=&\frac{1}{\kappa}\sqrt{\frac{3\alpha}{2}}
 \ln{\left\{1 + \frac{4n + \sqrt{16n^2 + 24\alpha n\left(1-n_s\right)\left(1+n\right)}}{3\alpha\left(1-n_s\right)}\right\}}~,\nonumber\\
 \Phi_\mathrm{f}&=&\frac{1}{\kappa}\sqrt{\frac{3\alpha}{2}}
 \ln{\left\{1 + \frac{2n}{\sqrt{3\alpha}}\right\}}
\end{eqnarray}
respectively, with $n_s$ being the spectral index of curvature perturbation. Moreover the Hubble parameter at the beginning of inflation (i.e $H_0$), 
in terms of $\alpha$, $n$ and $n_s$, is given by \cite{Drewes:2017fmn},
\begin{eqnarray}
 H_0 = \frac{8n\pi\sqrt{A_s}}{\kappa\sqrt{6\alpha}\left(e^{\kappa\Phi_\mathrm{i}\sqrt{\frac{2}{3\alpha}}} - 1\right)}~~.
 \label{Hubble-beginning}
\end{eqnarray}
Here $A_s$ denotes the amplitude of the curvature perturbation. Recall that the latest Planck 2018 data \cite{Planck:2018jri} puts a constraint 
on the scalar spectral index as $n_s = [0.9649 \pm 0.0042]$. 
Throughout the paper, we will consider $\alpha = 1$, $n = 1$, $n_s = 0.9625$ and $A_s = 2.2\times10^{-9}$, for which one gets 
$N_\mathrm{f} = 51.27$ and $H_0 = 1.6\times10^{13}\mathrm{GeV}$.

Eq.(\ref{cosmic Hubble parameter}) indicates that at the beginning 
of inflation, the cosmic Hubble parameter acquires $H_0 = 1/\eta_0$. Plugging the above expressions of 
$R$ and $\mathcal{G}$ into Eq.(\ref{form of f 1}) and a little bit of simplification yield the form of $f(R,\mathcal{G})$ as,
\begin{eqnarray}
 f(R,\mathcal{G}) = \kappa^{2q}\bigg\{\frac{\big[6\beta(\beta+1)\big]^q 
 + \big[-24(\beta+1)^3\big]^{q/2}}{\eta_0^{2q}}\bigg\}\bigg(\frac{-\eta}{\eta_0}\bigg)^{2\epsilon q}~~.
 \label{form of f 2}
\end{eqnarray}
It may be observed from the above expression that the condition $\epsilon = 0$ leads to the function $f(R,\mathcal{G})$ as a constant, 
for which the EM field Eq.(\ref{eom2}) becomes similar to the standard Maxwell's equation. This is however expected because for constant 
$f(R,\mathcal{G})$, the conformal breaking term 
$\mathcal{L}_{CB} = \sqrt{-g}\lambda f(R,\mathcal{G})~\epsilon^{\nu\nu\alpha\beta}F_{\mu\nu}F_{\alpha\beta}$ 
present in the action becomes a total surface term and thus gives no contributions in the field equations. As a result, the EM action preserves the 
conformal symmetry and thus the EM energy density 
decays by $1/a^4$ with the cosmological expansion of the universe, which in turn results to a very feeble magnetic amplitude at the current epoch. 
Thus the present model where the EM field couples with the background spacetime curvature via the term like 
$\sim \sqrt{-g}\lambda f(R,\mathcal{G})~\epsilon^{\nu\nu\alpha\beta}F_{\mu\nu}F_{\alpha\beta}$, requires 
$\epsilon \neq 0$ in order to break the conformal invariance of the EM field. 
Hence in order to generate sufficient strength of magnetic field and consequently the baryon asymmetry of the present universe in such higher 
curvature helical magnetogenesis set-up, we consider the background inflationary scenario as a quasi-de-Sitter one, in which case 
$\epsilon \neq 0$ and $\epsilon < 1$, in the subsequent calculations.

\section{Electric, magnetic and helicity power spectra}\label{sec_general expression}
 In this section we aim to calculate the energy-momentum tensor of the EM field and subsequently we will determine the individual energy density 
 of electric and magnetic field. In this regard, it may be mentioned that the electric and magnetic field are frame dependent, and for this 
 purpose, here we consider the $comoving~observer$, in which case the four velocity is given by $u^{\mu} = \left(a^{-1}(\eta),0,0,0\right)$ 
 in the $\left(\eta,\vec{x}\right)$ coordinate system. The calculation of electric, magnetic energy density at time $\eta$ 
 requires the energy-momentum tensor and the state (at $\eta$) of the EM field respectively. We will consider that the EM field 
 starts from the Bunch-Davies vacuum state at distant past (i.e the ``no-particle'' state), 
 which is indeed compatible with the equation of motion, as we will 
 show in the later section. Moreover, since we work in the Heisenberg picture, the state of the field remains fixed over time; however due to the 
 interaction between the EM field and the background FRW spacetime, the particle production occurs and thus the EM field does not remain in $vacuum$ 
 at later time. From action (\ref{action0}), we determine the energy-momentum tensor of the EM field as, 
\begin{eqnarray}
 T_{\alpha\beta}&=&-\frac{2}{\sqrt{-g}}~\frac{\delta}{\delta g^{\alpha\beta}}
 \left[\sqrt{-g}\left\{-\frac{1}{4}F_{\mu\nu}F^{\mu\nu} - \lambda~f(R,\mathcal{G})\epsilon^{\mu\nu\alpha\beta}F_{\mu\nu}F_{\alpha\beta}\right\}\right]\nonumber\\
 &=&-\frac{1}{4}\left\{g_{\alpha\beta}F_{\mu\nu}F^{\mu\nu} - 4g^{\mu\nu}F_{\mu\alpha}F_{\nu\beta}\right\} - \frac{2\lambda}{\sqrt{-g}}~
 \left[\mu\nu\rho\sigma\right]F_{\mu\nu}F_{\rho\sigma}\left(\frac{\delta f(R,\mathcal{G})}{\delta g^{\alpha\beta}}\right)~~,
 \label{em tensor}
\end{eqnarray}
where we use $\frac{\delta \sqrt{-g}}{\delta g^{\alpha\beta}} = -\frac{1}{2}\sqrt{-g}g_{\alpha\beta}$ and $\left[\nu\nu\rho\sigma\right]$ represents the 
antisymmetric permutation. It may be observed from Eq.(\ref{em tensor}) that $T_{\alpha\beta}$ 
contains an interaction energy density between electric and magnetic fields, which 
is generated entirely due to the presence of the parity violating term 
$\mathcal{L}_{CB} = \sqrt{-g}\lambda f(R,\mathcal{G})~\epsilon^{\nu\nu\alpha\beta}F_{\mu\nu}F_{\alpha\beta}$ in the action. As mentioned earlier, 
the parity violating term $\mathcal{L}_{CB}$ in the action leads to the helical nature of the magnetic field, in which case the helicity can 
be quantified by the quantity $\rho_h = -A_iB^{i}$, where $B^{i}$ specifies the magnetic field. At this stage, we proceed to quantize the EM field, 
i.e $A_i$ is promoted to a hermitian operator $\hat{A}_i$ as, 
\begin{eqnarray}
 \hat{A}_i(\eta,\vec{x}) = \int \frac{d\vec{k}}{(2\pi)^3}\sum_{r=+,-}\epsilon_{ri}~\bigg[\hat{b}_r(\vec{k})A_{r}(k,\eta)e^{i\vec{k}.\vec{x}} 
 + \hat{b}_r^{+}(\vec{k})A_{r}^{*}(k,\eta)e^{-i\vec{k}.\vec{x}}\bigg]~~,
 \label{mode decomposition}
\end{eqnarray}
where $\vec{k}$ is the Fourier mode momentum or equivalently the EM wave vector, $r$ is the polarization index, $\vec{\epsilon}_{+}$ and 
$\vec{\epsilon}_{-}$ are two polarization vectors, and $A_r(k,\eta)$ is the EM mode function corresponds to the mode $\vec{k}$. Due to the helical 
nature of the magnetic field, it will be useful if we work with the helicity basis set in the present context, in which case the 
polarization vectors are given by: $\vec{\epsilon}_{+} = \frac{1}{\sqrt{2}}\left(1,i,0\right)$ and 
$\vec{\epsilon}_{-} = \frac{1}{\sqrt{2}}\left(1,-i,0\right)$ respectively. The Coulomb gauge condition, i.e $\partial_{i}A^{i} = 0$ immediately 
leads to $\vec{k}.\vec{\epsilon}_{+} = \vec{k}.\vec{\epsilon}_{-} = 0$, which reveals that the EM propagation vector is perpendicular to the plane 
spanned by $\left(\vec{\epsilon}_{+},\vec{\epsilon}_{-}\right)$ vectors. The polarization vectors also satisfy the relation 
$\epsilon_{ijk}k_j\epsilon_{\pm k} = \mp ik~\epsilon_{\pm i}$ with $k = |\vec{k}|$, which will be useful later in solving the EM mode function. 
Furthermore $\hat{b}_r(\vec{k})$ and $\hat{b}_r^{+}(\vec{k})$ are the creation and annihilation operators of $\eta \rightarrow -\infty$ 
respectively, in particular they are defined over the vacuum state of distant past by the condition $\hat{b}_r(\vec{k})|0\rangle = 0$ $\forall$ $\vec{k}$. 
Using the Fourier mode decomposition of the EM field along with the commutation relation 
$\big[\hat{b}_p(\vec{k}),\hat{b}_r^{+}(\vec{k}')\big] = \delta_{pr}\delta\big(\vec{k} - \vec{k}'\big)$, the 
expectation values (over the Bunch-Davies vacuum state) of electric and magnetic energy densities come as,
\begin{eqnarray}
 \big\langle \rho(\vec{E}) \big\rangle&=&\sum_{r=+,-}\int \frac{dk}{2\pi^2}~\frac{k^2}{a^4}\big|A_r'(k,\eta)\big|^2\nonumber\\
 \big\langle \rho(\vec{B}) \big\rangle&=&\sum_{r=+,-}\int \frac{dk}{2\pi^2}~\frac{k^4}{a^4}\big|A_r(k,\eta)\big|^2
 \label{expectation energy density 1}
 \end{eqnarray}
respectively. Similarly, the helicity energy density in the Fourier basis is given by,
\begin{eqnarray}
 \big\langle \rho_h \big\rangle = \int \frac{dk}{2\pi^2}~\frac{k^3}{a^3}\left\{\big|A_{+}(k,\eta)\big|^2 - \big|A_{-}(k,\eta)\big|^2\right\}~~.
 \label{expectation energy density 3}
\end{eqnarray}
Consequently the power spectra of electric and magnetic fields (defined by the energy density within unit logarithmic interval of $k$) become, 
\begin{eqnarray}
 \mathcal{P}(\vec{E}) = \frac{\partial \big\langle \rho(\vec{E}) \big\rangle}{\partial \ln{k}} 
 = \sum_{r=+,-} \frac{k}{2\pi^2}~\frac{k^2}{a^4}\big|A_r'(k,\eta)\big|^2~~~~~~~~~~,~~~~~~~~~~~
 \mathcal{P}(\vec{B}) = \frac{\partial \big\langle \rho(\vec{B}) \big\rangle}{\partial \ln{k}} 
 = \sum_{r=+,-} \frac{k}{2\pi^2}~\frac{k^4}{a^4}\big|A_r(k,\eta)\big|^2~~.
 \label{power spectra}
\end{eqnarray}
Furthermore, the helicity power spectrum is given by,
\begin{eqnarray}
 \mathcal{P}_h = \frac{\partial \big\langle \rho_h \big\rangle}{\partial \ln{k}} 
 = \frac{k}{2\pi^2}~\frac{k^3}{a^3}\left\{\big|A_{+}(k,\eta)\big|^2 - \big|A_{-}(k,\eta)\big|^2\right\}~~.
 \label{helicity power spectrum}
\end{eqnarray}
Here it may be observed that the scale dependence of $\mathcal{P}(\vec{E})$, $\mathcal{P}(\vec{B})$ and $\mathcal{P}_h$ are different. In particular, 
the electric power spectrum gets the scale dependency from the $k^3$ factor as well as from the time derivative of the EM mode function, while the magnetic power 
spectrum has the scale dependency due to the $k^5$ factor and also due to the mode function itself. On the other hand, the 
scale dependency of the helicity spectrum comes through the $k^3$ factor and from the difference between the two EM mode functions. Therefore it is very 
unlikely, that the electric, magnetic and helicity power spectra have similar dependency on $k$. In order to extract the scale dependence of such 
power spectra, we need to solve the EM mode functions from Eq.(\ref{eom2}), which is the subject of the next section.

\section{Solution of the mode function and scale dependence of the power spectra during inflation}\label{sec_solution_inflation}
In this section we aim to determine the solution of EM mode functions and the corresponding power spectra during the inflationary era 
when the scale factor behaves as of Eq.(\ref{scale factor}). For this purpose, we need Eq.(\ref{eom2}) which, due to the Fourier decomposition of 
$A_i(\eta,\vec{x})$, takes the following form:
\begin{eqnarray}
 A_{\pm}''(k,\eta) + \left(k^2 \pm 8\lambda k~f'(R,\mathcal{G})\right)A_{\pm}(k,\eta) = 0~~,
 \label{FT eom1}
\end{eqnarray}
where $f'(R,\mathcal{G}) = \frac{df}{d\eta}$. The form of 
$f(R,\mathcal{G})$ of Eq.(\ref{form of f 1}) immediately leads to  
\begin{eqnarray}
 A_{\pm}''(k,\eta) + \left(k^2 \mp k\left(\frac{\zeta^2}{\eta^2}\right)\left(\frac{-\eta_0}{\eta}\right)^{2\alpha}\right)A_{\pm}(k,\eta) = 0~~,
 \label{FT eom3}
\end{eqnarray}
where $\zeta^2$ and $\alpha$ have the following forms,
\begin{eqnarray}
 \zeta^2&=&\left(16\epsilon q\lambda\eta_0\right)\left(\frac{\kappa}{\eta_0}\right)^{2q}\bigg\{\big[6\beta(\beta+1)\big]^q 
 + \big[-24(\beta+1)^3\big]^{q/2}\bigg\}~,\nonumber\\
 \alpha&=&-\frac{1}{2} - \epsilon q
 \label{B}
\end{eqnarray}
Here it may be mentioned that $\zeta^2$ is proportional with $\epsilon$, $q$ and thus for $\epsilon = 0$ or $q = 0$, both the 
EM mode functions seem to satisfy the standard Maxwell's equation in vacuum, as expected. Thereby the non-zero values of 
$\epsilon$ (the slow roll parameter) and $q$ modify the EM field equations non-trivially by Eq.(\ref{FT eom3}), compared to the standard 
Maxwell's equation. 

Eq.(\ref{FT eom3}) may not be solved in a closed form. To obtain 
the solution, we individually consider the sub-horizon (region I) and super-horizon (region II) limits respectively. 
In Region I (sub-horizon limit), the wavelength of the 
mode is smaller than the Hubble radius, i. e. $H \ll k$, and thus one can neglect the term containing $\zeta^2$ in Eq.(\ref{FT eom3}). 
In Region II (super-horizon scales), the mode lies outside the Hubble radius, i. e. $k \ll H$, and thus we 
can neglect $k^2$ in Eq.(\ref{FT eom3}). While evaluating
the mode-functions is trivial in Region I, it is highly non-trivial in Region II.

In region I, Eq.(\ref{FT eom3}) can be expressed as,
\begin{eqnarray}
 A_{\pm}''(k,\eta) + k^2A_{\pm}(k,\eta) = 0~.
 \label{FT eom region I}
\end{eqnarray}
The Bunch-Davies vacuum state is considered to be the initial states of the EM mode functions, due to which, the solution of Eq.(\ref{FT eom region I}) is
\begin{eqnarray}
 A_{\pm}(k,\eta) = \frac{1}{\sqrt{2k}}e^{-ik\eta}~.
 \label{FT solution region I}
\end{eqnarray}
In region II (super-horizon scale), Eq.(\ref{FT eom3}) turns out to be,
\begin{eqnarray}
 A_{\pm}''(k,\eta) \mp k\left(\frac{\zeta^2}{\eta^2}\right)\left(\frac{-\eta_0}{\eta}\right)^{2\alpha}A_{\pm}(k,\eta) = 0~~,
 \label{FT eom region II}
\end{eqnarray}
on solving which, we get $A_{\pm}(k,\eta)$ as (see the Appendix-A in Sec.[\ref{sec-app1}]),
\begin{eqnarray}
 A_{+}(k,\eta)&=&\left(\frac{-\eta}{\eta_0}\right)^{1/2}\left\{C_1~
 J_{\frac{1}{2\alpha}}\left(-i\frac{\zeta\sqrt{k}}{\alpha}\left(\frac{-\eta_0}{\eta}\right)^{\alpha}\right) 
 + C_2~Y_{\frac{1}{2\alpha}}\left(-i\frac{\zeta\sqrt{k}}{\alpha}\left(\frac{-\eta_0}{\eta}\right)^{\alpha}\right)\right\}~~,\nonumber\\
 A_{-}(k,\eta)&=&\left(\frac{-\eta}{\eta_0}\right)^{1/2}\left\{C_3~
 J_{\frac{1}{2\alpha}}\left(\frac{\zeta\sqrt{k}}{\alpha}\left(\frac{-\eta_0}{\eta}\right)^{\alpha}\right) 
 + C_4~Y_{\frac{1}{2\alpha}}\left(\frac{\zeta\sqrt{k}}{\alpha}\left(\frac{-\eta_0}{\eta}\right)^{\alpha}\right)\right\}~~,
 \label{sol1}
\end{eqnarray}
where $J_{\frac{1}{2\alpha}}(z)$ and $Y_{\frac{1}{2\alpha}}(z)$ represent the Bessel functions of first and second kind respectively, and 
$C_i$ ($i = 1,2,3,4$) are the integration constants. The integration constants can be determined by matching $A_{\pm}(k,\eta)$ and 
$A_{\pm}'(k,\eta)$ at the transition time between the regions I and II, i.e when the mode $k$ crosses the horizon. If the horizon crossing instant 
of $k$-th mode is symbolized by $\eta_{*}$, then we can write,
\begin{eqnarray}
 \left|k\eta_{*}\right| = 1+\epsilon~.
 \label{horizon crossing condition}
\end{eqnarray}
The matching conditions of $A_{\pm}(k,\eta)$ and $A_{\pm}'(k,\eta)$ at $\eta_{*}$ lead to the explicit expressions of $C_i$ ($i=1,2,3,4$) which we present 
in the Appendix-A in Sec.[\ref{sec-app1}]. The above expressions of $A_{\pm}(k,\eta)$ 
immediately yields the electric and magnetic power spectra in the superhorizon regime as (see the Appendix-A in Sec.[\ref{sec-app1}] for 
detailed calculations),
\begin{eqnarray}
 \mathcal{P}(\vec{E}) = \left(\frac{k}{2\pi^4}\right)\left(\frac{H_0}{k}\right)^2\left(\frac{k}{a}\right)^4
 \left|\left(\frac{\zeta\sqrt{k}}{2\alpha}\right)^{-\frac{1}{2\alpha}}\Gamma\left(\frac{1}{2\alpha}\right)\right|^2\left\{\left|C_2\right|^2 + \left|C_4\right|^2\right\}
 \label{electric power spectrum 1}
\end{eqnarray}
and
 \begin{eqnarray}
 \mathcal{P}(\vec{B}) = \left(\frac{k}{2\pi^2}\right)\left(\frac{k}{a}\right)^4\left|\frac{\left(\frac{\zeta\sqrt{k}}{2\alpha}\right)^{\frac{1}{2\alpha}}}
 {\Gamma\left(1 + \frac{1}{2\alpha}\right)}\right|^2
 \left\{\left|C_1 - C_2\cot{\left(-\frac{\pi}{2\alpha}\right)}\right|^2 + 
 \left|C_3 - C_4\cot{\left(-\frac{\pi}{2\alpha}\right)}\right|^2\right\}
 \label{magnetic power spectrum 1}
\end{eqnarray}
respectively. Eqs.(\ref{electric power spectrum 1}) and (\ref{magnetic power spectrum 1}) 
clearly demonstrate that the electric and magnetic power spectra are 
not scale invariant during the superhorizon inflationary era, however they have different scale dependence. Furthermore both the 
$\mathcal{P}(\vec{E})$ and $\mathcal{P}(\vec{B})$ tends to zero as 
$|k\eta| \rightarrow 0$. Such characteristics of the power spectra hints to the resolution of the 
backreaction problem in the present magnetogenesis model. 

\begin{figure}[!h]
\begin{center}
 \centering
 \includegraphics[width=3.5in,height=2.5in]{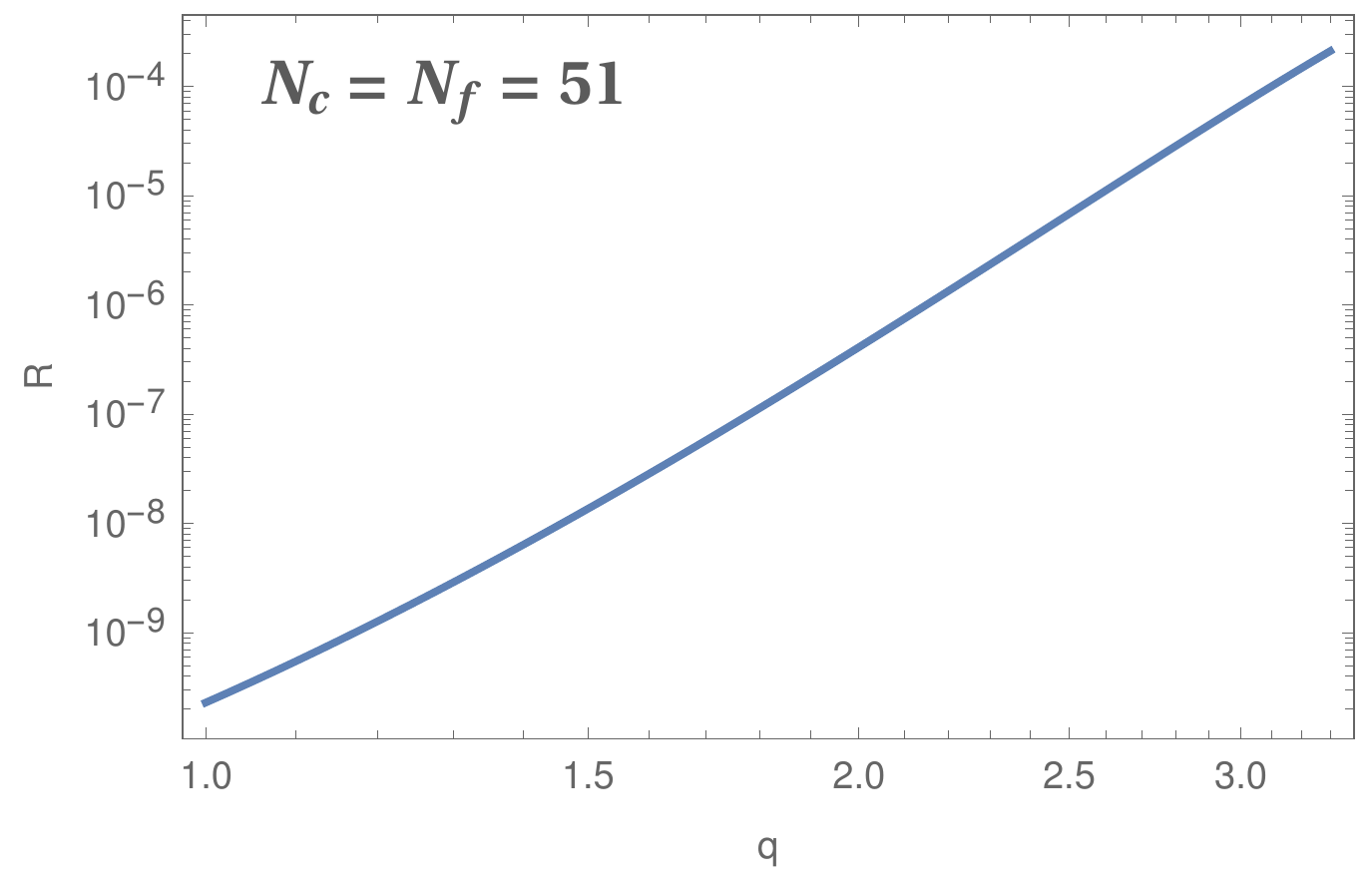}
 \includegraphics[width=3.5in,height=2.5in]{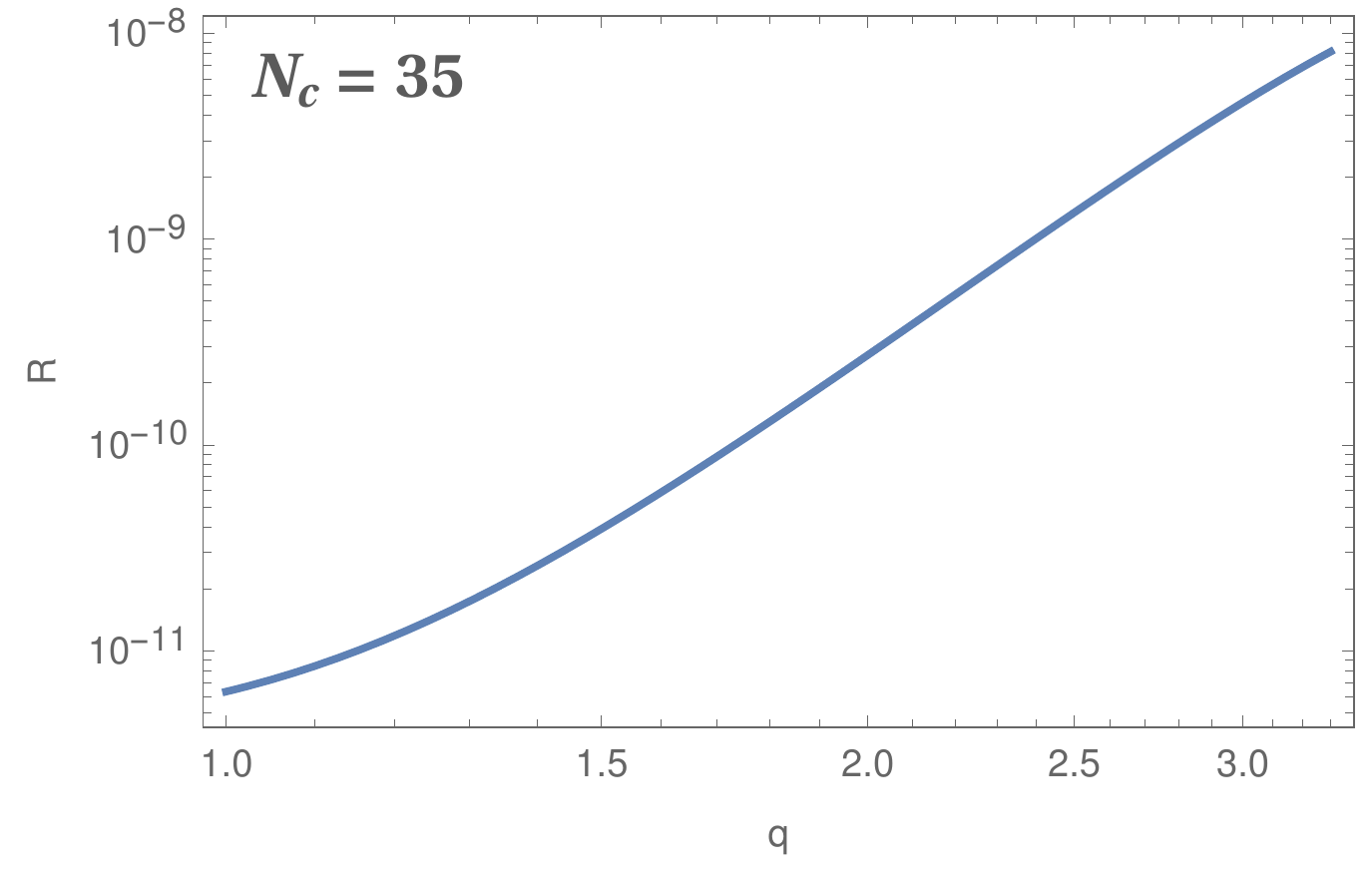}
 \caption{$R$ vs $q$ for $H_0 = 1.6\times10^{13}\mathrm{GeV}$, $N_f = 51$ and $\epsilon = 0.1$. Moreover we take $\lambda = 1$. In the left plot, 
 $k_c = 10^{-20}\mathrm{GeV}$ that crosses the horizon near the end of inflation, i.e $N_c = N_f = 51$; while for the right plot, 
 $k_c = 10^{-25}\mathrm{GeV}$ which crosses the horizon near $N_c = 35$ measured from the beginning of inflation.}
 \label{plot-backreaction}
\end{center}
\end{figure}

However in order to ensure that the backreaction of the EM field on the background spacetime stays small during the inflation, 
we first consider the electric field and magnetc field energy density at $\eta = \eta_c$, which are given by,
\begin{eqnarray}
 \rho(\vec{E},\eta_c)=\int_{k_i}^{k_c} \mathcal{P}(\vec{E})~d\ln{k}~~~~~~~~~~\mathrm{and}~~~~~~~~~~~
 \rho(\vec{B},\eta_c) = \int_{k_i}^{k_c} \mathcal{P}(\vec{B})~d\ln{k}~~.
 \label{electric energy density 1}
\end{eqnarray}
The integration limits $k_i$ and $k_c$ are the mode momentum which cross the horizon at the beginning of inflation and at $\eta = \eta_c$ respectively. 
The beginning of inflation is considered to be the instance when the CMB scale mode crosses the horizon, and 
thus $k_i = k_\mathrm{CMB} = 0.05\mathrm{Mpc}^{-1}$. Moreover we have $|\eta_c| = k_c^{-1}$ where $\eta_c$ is any intermediate time 
instance during the inflation and thus $k_c$ is greater than the CMB scale momentum. The quantity $N_c$ represents the inflationary 
e-folding number up-to $\eta = \eta_c$ measured from the beginning of inflation, 
i.e $N_c = \ln{\left(\frac{a_c}{a_i}\right)}$ with $a_c = a(\eta_c)$ and $a_i = a(\eta_i)$. In 
order to resolve the backreaction issue, we need to examine whether the EM energy density is lower than the 
background energy density which is given by $\rho_\mathrm{bg} = 3H^2M_\mathrm{Pl}^2$, where $M_\mathrm{Pl}$ is the reduced Planck mass. 
However due to the dependence of $C_i = C_i(k)$ ($i=1,2,3,4$) (see the Sec.[\ref{sec-app1}]), the integration in Eq.(\ref{electric energy density 1}) 
may not be performed in a closed form, and thus we numerically integrate Eq.(\ref{electric energy density 1}) to obtain the electric and magnetic 
energy density at $\eta_c$. For this purpose, we consider $H_0 = 1.6\times10^{13}\mathrm{GeV}$, $N_f = 51$ and $\epsilon =  0.1$ (see the discussion after 
Eq.(\ref{Hubble-beginning})). Considering these set of values, we perform the numerical integration of Eq.(\ref{electric energy density 1}), and 
give a plot of the quantity 
\begin{eqnarray}
 R = \frac{\left(\rho(\vec{E},\eta_c) + \rho(\vec{B},\eta_c)\right)}{\rho_{bg}}~~,\nonumber
\end{eqnarray}
with respect to the parameter $q$ for different values of $k_c$, see Fig.[\ref{plot-backreaction}]. 
In particular, we consider $k_c = 10^{-20}\mathrm{GeV}$ and $10^{-25}\mathrm{GeV}$ in the left and 
right plot of Fig.[\ref{plot-backreaction}] respectively. Here it deserves mentioning that the mode 
$k_c = 10^{-20}\mathrm{GeV}$ crosses the horizon near the end of inflation, i.e $N_c \approx N_f$; while the mode $k_c = 10^{-25}\mathrm{GeV}$ crosses 
the horizon near $N_c = 35$ measured from the beginning of inflation. 

The Fig.[\ref{plot-backreaction}] clearly demonstrates that $R \leq 10^{-4}$ for 
$q \leq 3.3$. This argues that for $q \leq 3.3$, the EM field has negligible backreaction 
on the background inflationary spacetime -- leading to the resolution of the backreaction problem in the present context.\\  

In regard to the helicity of the magnetic field, Eq.(\ref{helicity power spectrum}) leads to the helicity power spectrum during the inflation as,
\begin{eqnarray}
 \mathcal{P}_h 
 = \left(\frac{k}{2\pi^2}\right)\left(\frac{k}{a}\right)^3\left|\frac{\left(\frac{\zeta\sqrt{k}}{2\alpha}\right)^{1/(2\alpha)}}
 {\Gamma\left(1 + \frac{1}{2\alpha}\right)}\right|^2
 \left\{\left|C_1 - C_2\cot{\left(-\frac{\pi}{2\alpha}\right)}\right|^2 - 
 \left|C_3 - C_4\cot{\left(-\frac{\pi}{2\alpha}\right)}\right|^2\right\}~~.
 \label{helicity power spectrum inflation}
\end{eqnarray}
Eq.(\ref{helicity power spectrum inflation}) reveals that the comoving helicity remains conserved with the expansion of the universe 
during the inflationary super-Hubble regime. Moreover it is evident from Eq.(\ref{sol1}) that $A_{+}(k,\eta)$ contains a complex argument 
within the Bessel function, which in turn makes the amplitude of the positive helicity mode much larger than that of the negative helicity mode, 
i.e $\left|A_{+}(k,\eta)\right| \gg \left|A_{-}(k,\eta)\right|$. Thus 
the helicity power spectrum of the EM field turns out to be positive.\\

\section{Present magnetic strength and magnetic helicity for the instantaneous reheating case}\label{sec_instantaneous reheating}
\begin{figure}[!h]
\begin{center}
 \centering
 \includegraphics[width=3.5in,height=2.5in]{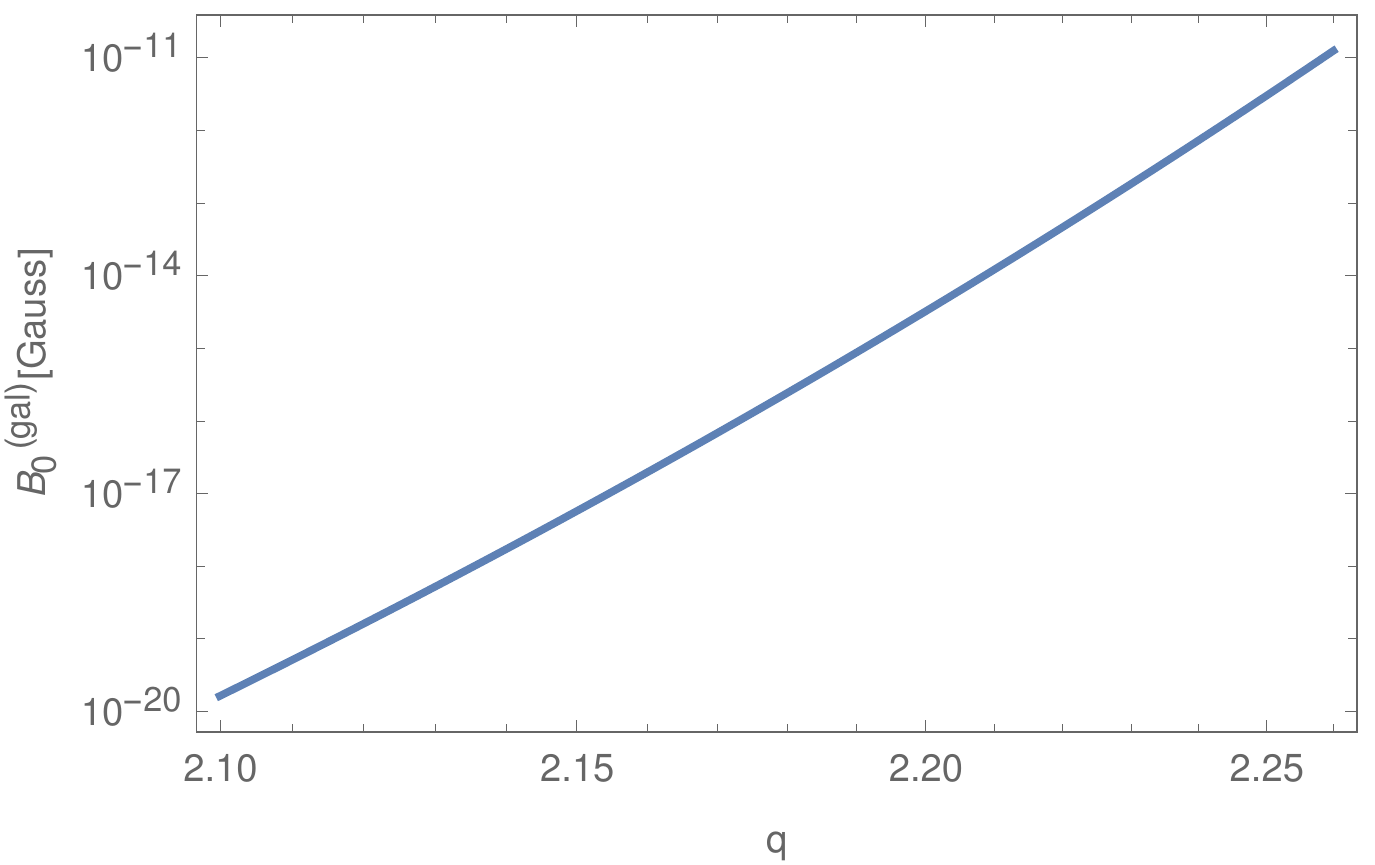}
 \caption{$B_0^\mathrm{(gal)}$ vs $q$ from Eq.(\ref{magnetic strength 2}). We take $\lambda = 1$.}
 \label{plot-present-mag-str}
\end{center}
\end{figure}

In this section, we aim to calculate the magnetic strength and consequently the magnetic helicity at current universe. For this purpose, 
we need to know about the conductivity of the universe during and after the inflation. During the inflation, the universe remains a poor electrical 
conductor where the conductivity is low, however in the post-inflationary era, the universe becomes a good conductor. The large conductivity 
during the post-inflation lies in the consideration that the universe experiences an $instantaneous~reheating$ phase just after the inflation, 
in which case, the e-fold number of the reheating epoch is zero and thus the universe makes a sudden jump from the inflation 
to a radiation dominated epoch where the Hubble parameter evolves as $H \propto t^{-1}$ with $t$ being the cosmic time (here 
we would like to mention that at a later section, we will relax the assumption of ``instantaneous reheating'' and will 
consider a reheating epoch having non-zero e-fold number after inflation). The high conductivity in the post-inflation phase leads to an 
electric current of the form $J^{i} = \sigma E^{i}$ (where $\sigma$ is the electrical conductivity and $E^{i}$ is the electric field) 
which in turn acts as a source term in the EM field equation. Due to the presence of such source term and owing to the large $\sigma$, the solution 
of the EM field after inflation behaves as $A_{\pm}(k,\eta) = C_{\pm} + D_{\pm}e^{-\sigma t}$ (with $C_{\pm}$ and $D_{\pm}$ being the 
integration constants), where the term $e^{-\sigma t}$ becomes soon negligible and thus $A_{\pm}(k,\eta)$ becomes constant with time. As a 
result, the electric field becomes vanishingly small and the magnetic field remains the dominant piece in the electromagnetic energy density. 
Moreover we mentioned earlier that at the late time, the conformal breaking coupling $f(R,\mathcal{G})$ does not contribute and thus the EM action 
restores the conformal symmetry of the EM field. In particular, the function $f(R,\mathcal{G})$ is considered to be zero from the end of inflation, 
which is also connected from the continuity point of view, as Eq.(\ref{form of f 2}) leads to $f(R,\mathcal{G})$ tends to zero as $|k\eta| \rightarrow 0$, 
i.e at the end of inflation. From the conformal symmetry of the EM field, we may argue that the EM energy density decays as $\frac{1}{a^4}$ 
with the expansion of the universe, or equivalently the magnetic energy density goes by $1/a^4$, as the electric field practically vanishes 
in the post-inflationary epoch. As a result, the magnetic energy density at the end of inflation is related to that of at the 
present epoch by the following relation,
\begin{eqnarray}
 \mathcal{P}(\vec{B})\bigg|_{0} = \bigg(\frac{a_f}{a_0}\bigg)^4~\mathcal{P}(\vec{B})\bigg|_{f}~~,
 \label{magnetic strength 1}
\end{eqnarray}
where the suffix 'f' and '0' denote the end instance of inflation and the current time of universe respectively. Moreover 
the helicity power spectrum evolves as $1/a^3$ in the post-inflationary era, and 
thus $\mathcal{P}_h$ at the end of inflation is connected to that of at the present epoch by following:
\begin{eqnarray}
 \mathcal{P}_h\bigg|_{0} = \bigg(\frac{a_f}{a_0}\bigg)^3~\mathcal{P}_h\bigg|_{{f}}~~.
 \label{helicity strength 1}
\end{eqnarray}
Eqs.(\ref{magnetic power spectrum 1}) and (\ref{magnetic strength 1}) lead to the magnetic field's current amplitude as,
\begin{eqnarray}
 B_0 = \left(\frac{k}{\pi^2}\right)^{1/2}\left|\frac{\left(\frac{\zeta\sqrt{k}}{2\alpha}\right)^{1/(2\alpha)}}
 {\Gamma\left(1 + \frac{1}{2\alpha}\right)}\right|
 \left|C_1 - C_2\cot{\left(-\frac{\pi}{2\alpha}\right)}\right|\left(\frac{a_f}{a_0}\right)^2\left(\frac{k}{a_f}\right)^2~,
 \label{magnetic strength 2}
\end{eqnarray}
with, recall that $C_i$ ($i=1,2$) are given in the Sec.[\ref{sec-app1}] and $k = 0.05\mathrm{Mpc}^{-1}$ 
lies around the large scale CMB mode. Here we consider the contribution from the positive helicity mode only, due to the reason that 
$\left|A_{+}(k,\eta)\right| \gg \left|A_{-}(k,\eta)\right|$ during the inflationary era. If this generated magnetic fields on 
 the scale $k = 0.05\mathrm{Mpc}^{-1}$ at the present time play the role of seed magnetic fields of galactic magnetic fields, then by assuming 
 the magnetic flux conservation, $Br^2 = \mathrm{constant}$ \cite{Bamba:2006km}, where $r$ is a scale, and using the scale 
 ratio $kr_\mathrm{gal}/\left(2\pi\right) \sim 10^{-5}$ , where $r_\mathrm{gal}$ is the scale of galaxies, we can estimate the 
 strength of the magnetic fields at the galactic scale as,
 \begin{eqnarray}
  B_0^\mathrm{(gal)} = \left(\frac{2\pi}{k r_\mathrm{gal}}\right)^2
  \left(\frac{k}{\pi^2}\right)^{1/2}\left|\frac{\left(\frac{\zeta\sqrt{k}}{2\alpha}\right)^{1/(2\alpha)}}
 {\Gamma\left(1 + \frac{1}{2\alpha}\right)}\right|
 \left|C_1 - C_2\cot{\left(-\frac{\pi}{2\alpha}\right)}\right|\left(\frac{a_f}{a_0}\right)^2\left(\frac{k}{a_f}\right)^2~~.
  \label{galactic magnetic field}
 \end{eqnarray}
Here $a_f \approx \left(-H_f~\eta_f\right)^{-1}$. Therefore 
in order to estimate $B_0^\mathrm{(gal)}$ from Eq.(\ref{galactic magnetic field}), we need to know $\frac{a_f}{a_0}$ and 
$|k\eta_f|$. In regard to $\frac{a_f}{a_0}$, we use 
 the entropy conservation in the post-inflation era, i.e $gT^3a^3 = \mathrm{constant}$, where $g$ symbolizes the effective relativistic 
 degrees of freedom and 
 $T$ is the corresponding temperature -- this finally results to $\frac{a_0}{a_f} = 10^{30}\left(H_f/10^{-5}M_\mathrm{Pl}\right)^{1/2}$, 
 where $H_f$ is the 
 Hubble parameter at the end of inflation and can be determined from Eq.(\ref{cosmic Hubble parameter}) as 
 $H_f = H_0\exp{\left(-\epsilon N_\mathrm{f}\right)}$, with $N_\mathrm{f}$ being the total e-fold of the inflationary phase. 
 The set: $H_0 = 1.6\times10^{13}\mathrm{GeV}$, $\epsilon = 0.1$ and $N_f = 51$ immediately 
 leads to $H_f = 9\times10^{10}\mathrm{GeV}$. Moreover, for the purpose of determining $|k\eta_f|$, the relation $|\eta_f| = 1/k_f$ 
 stands to be useful where $k_f$ is the momentum of the mode which crosses the Hubble horizon at the end instance of inflation, and 
 $k$ lies around the CMB scale. As a consequence, $|k\eta_f|$ is given by 
 $|k\eta_f| = \exp{\left(-N_\mathrm{f}\right)}$ where $k = 0.05\mathrm{Mpc}^{-1} = 3.2\times10^{-40}\mathrm{GeV}$. 
 Using such expressions and considering a sample value of $q = 2.18$, we get,
 \begin{eqnarray}
  \left|\frac{\left(\frac{\zeta\sqrt{k}}{2\alpha}\right)^{1/(2\alpha)}}
 {\Gamma\left(1 + \frac{1}{2\alpha}\right)}\right|
 \left|C_1 - C_2\cot{\left(-\frac{\pi}{2\alpha}\right)}\right| \sim 10^{54}~~.\nonumber
 \end{eqnarray}
Consequently, for $q = 2.18$, the $B_0^\mathrm{(gal)}$ is estimated as,
\begin{eqnarray}
 B_0^\mathrm{(gal)} \sim 10^{-16}\mathrm{Gauss}~~,\nonumber
\end{eqnarray}
where the conversion $1\mathrm{G} = 1.95\times10^{-20}\mathrm{GeV}^2$ is used. 
For a better understanding, 
we give a plot of $B_0^\mathrm{(gal)}$ vs. $q$ from Eq.(\ref{galactic magnetic field}) in Fig.[\ref{plot-present-mag-str}].\\ 
 
In order to confront the model with observations, we use the CMB results, 
according to which, the present magnetic field around the CMB scale lies within 
$10^{-22}\mathrm{G} \lesssim B_0^\mathrm{(gal)} \lesssim 10^{-10}\mathrm{G}$. 
Thereby the Fig.[\ref{plot-present-mag-str}] demonstrates that the theoretically predicted $B_0^\mathrm{(gal)}$ gets consistent with 
the observational constraints for the parametric regime: $2.1 \leq q \leq 2.26$. Here we need to recall that $q$ should lie within $q < 3.3$ 
to have a negligible backreaction of the EM field on the background spacetime (see Fig.[\ref{plot-backreaction}]). 
Thus we may argue that the magnetogenesis model, where the EM field gets non-minimally coupled by the term 
$\mathcal{L}_{CB} = \sqrt{-g}\lambda f(R,\mathcal{G})~\epsilon^{\nu\nu\alpha\beta}F_{\mu\nu}F_{\alpha\beta}$, seems to generate 
sufficient magnetic strength at present epoch of the universe and concomitantly resolves the backreaction issue.\\

In regard to the generation of BAU from helical magnetic fields, the net baryonic number density ($n_\mathrm{B}$) is defined by 
\cite{Bamba:2006km,Giovannini:1997eg},
\begin{eqnarray}
 n_\mathrm{B} = -\frac{n_\mathrm{f}}{2}\Delta n_\mathrm{CS}~~~~~~~~~~,~~~~~~~~~~~\Delta n_\mathrm{CS} = -\frac{g'^2}{4\pi^2}\int^{t}\vec{E}.\vec{B}~dt~~.
 \label{definition1}
\end{eqnarray}
Here, $n_\mathrm{f}$ 
is the number of fermionic generations (throughout this paper we use $n_\mathrm{f}=3$), 
$\Delta n_\mathrm{CS}$ is the Chern-Simons number density and $g'^2/\left(4\pi\right) = \alpha_\mathrm{EM}$ is the fine structure constant. 
The electric and magnetic fields are defined by \cite{Bamba:2006km},
\begin{eqnarray}
 E_{i} = -\frac{1}{a}\dot{A}_{i}~~~~~~~~~~\mathrm{and}~~~~~~~~~~~B_{i} = \frac{1}{a^2}\epsilon_{ijk}\partial_{j}A_{k}~~.
\end{eqnarray}
Owing to the mode decomposition of $A_{i}(\vec{x},\eta)$, $n_\mathrm{B}$ can be determined as,
\begin{eqnarray}
 n_\mathrm{B}&=&\left(\frac{n_\mathrm{f}}{2}\right)\left(\frac{g'^2}{4\pi^2}\right)\mathcal{P}_h\bigg|_{0}\nonumber\\
 &=&\left(\frac{n_\mathrm{f}}{2}\right)\left(\frac{g'^2}{4\pi^4}\right)\left(\frac{k}{2\pi^2}\right)\left(\frac{k_0}{a_{0}}\right)^3
 \left|\frac{\left(\frac{\zeta\sqrt{k}}{2\alpha}\right)^{1/(2\alpha)}}
 {\Gamma\left(1 + \frac{1}{2\alpha}\right)}\right|^2\left|C_1 - C_2\cot{\left(-\frac{\pi}{2\alpha}\right)}\right|^2~~.
 \nonumber
 \label{baryon number-1}
\end{eqnarray}
Here $k_0 = a_0H_0$, i.e $k_0^{-1}$ is the present horizon scale. Due to the 
Eq.(\ref{galactic magnetic field}), the above expression 
of $n_\mathrm{B}$ can be equivalently written as \cite{Bamba:2006km},
\begin{eqnarray}
 n_\mathrm{B} = \left(\frac{n_\mathrm{f}}{2}\right)\left(\frac{g'^2}{4\pi^2}\right)\left(\frac{a_0}{k_0}\right)
 \left(\frac{k r_\mathrm{gal}}{2\pi}\right)^4\left(B_0^\mathrm{(gal)}\right)^2~~.
 \label{baryon number-2}
\end{eqnarray}
In regard to estimate the net baryon density, we will use Eq.(\ref{baryon number-2}) where $n_\mathrm{B}$ has been expressed in terms of 
the $B_0^\mathrm{(gal)}$. However it is more convenient to get the ratio of net baryon density to the entropy density of the present universe,
i.e $n_\mathrm{B}/s_0$, 
which is dimensionless and has observational constraints as $n_\mathrm{B}/s_0 = \left(6.09 \pm 0.06\right)\times10^{-10}$ from the CMB data. The current 
entropy density ($s_0$) is given by,
\begin{eqnarray}
 s_0 = 2.97\times10^{3}\left(\frac{T_0}{2.75\left[K\right]}\right)^3~\mathrm{cm}^{-3}~~,
 \label{entropy}
\end{eqnarray}
with $T_0 = 2.73K$ is the present temperature of the cosmic microwave background (CMB) radiation. Fig.[\ref{plot2}] depicts that the model 
predicts sufficient magnetic strength for a suitable regime of the reheating EoS parameter. Using this information along with 
the parameter values that we considered earlier (i.e $H_0 = 1.6\times10^{13}\mathrm{GeV}$, $\epsilon = 0.1$, $N_\mathrm{f} = 51$ and 
$H_f = 9\times10^{10}\mathrm{GeV}$), we estimate $n_\mathrm{B}/s_0$ for 
different values of $B_0^{\mathrm{gal}}$, with $\frac{k_0}{a_0} = 0.05\mathrm{Mpc}^{-1} \approx 3.2\times10^{-40}\mathrm{GeV}$. 
These are given in Table[\ref{Table-1}]. From Table[\ref{Table-1}], we see that the magnetic fields of current strength 
$B_0^\mathrm{(gal)} \sim 10^{-13}\mathrm{G}$ leads to the resultant value of $n_\mathrm{B}/s_0$ as of the order $\sim 10^{-10}$.  
 \begin{table}[h]
  \centering
 \resizebox{\columnwidth}{2.0 cm}{%
  \begin{tabular}{|c|c|}
   \hline 
    Current magnetic field: $B_0^{\mathrm{gal}}\left[\mathrm{Gauss}\right]$ & $n_\mathrm{B}/s_0$ (Dimensionless)\\
   \hline
   $3.27\times10^{-13}$ & $6.09\times10^{-10}$ \\
   \hline
   $3.26\times10^{-13}$ & $6.05\times10^{-10}$ \\
   \hline
   $3.28\times10^{-13}$ & $6.12\times10^{-10}$ \\
   \hline
   $3.29\times10^{-13}$ & $6.14\times10^{-10}$ \\
   \hline
   \hline
  \end{tabular}%
 }
  \caption{The ratio of net baryon density to the present entropy density (i.e $n_\mathrm{B}/s_0$) for different magnetic strength $B_0^\mathrm{(gal)}$ 
  with $\frac{k_0}{a_0} = 0.05\mathrm{Mpc}^{-1} \approx 3.2\times10^{-40}\mathrm{GeV}$.}
  \label{Table-1}
 \end{table}

Here we would like to mention that the above arguments are based on $instantaneous~reheating$ where, 
as mentioned earlier, the reheating epoch has zero e-fold and thus the 
universe gets a large conductivity after the end of inflation. However it would be more physical if the universe undergoes through a reheating phase 
having a non-zero e-fold number. Motivated by this, in the next section, we will relax the assumption of ``instantaneous reheating'' 
and will consider a prolonged reheating stage in-between the end of inflation and the beginning of radiation dominated epoch.\\

\begin{figure}[!h]
\begin{center}
 \centering
 \includegraphics[width=3.5in,height=2.5in]{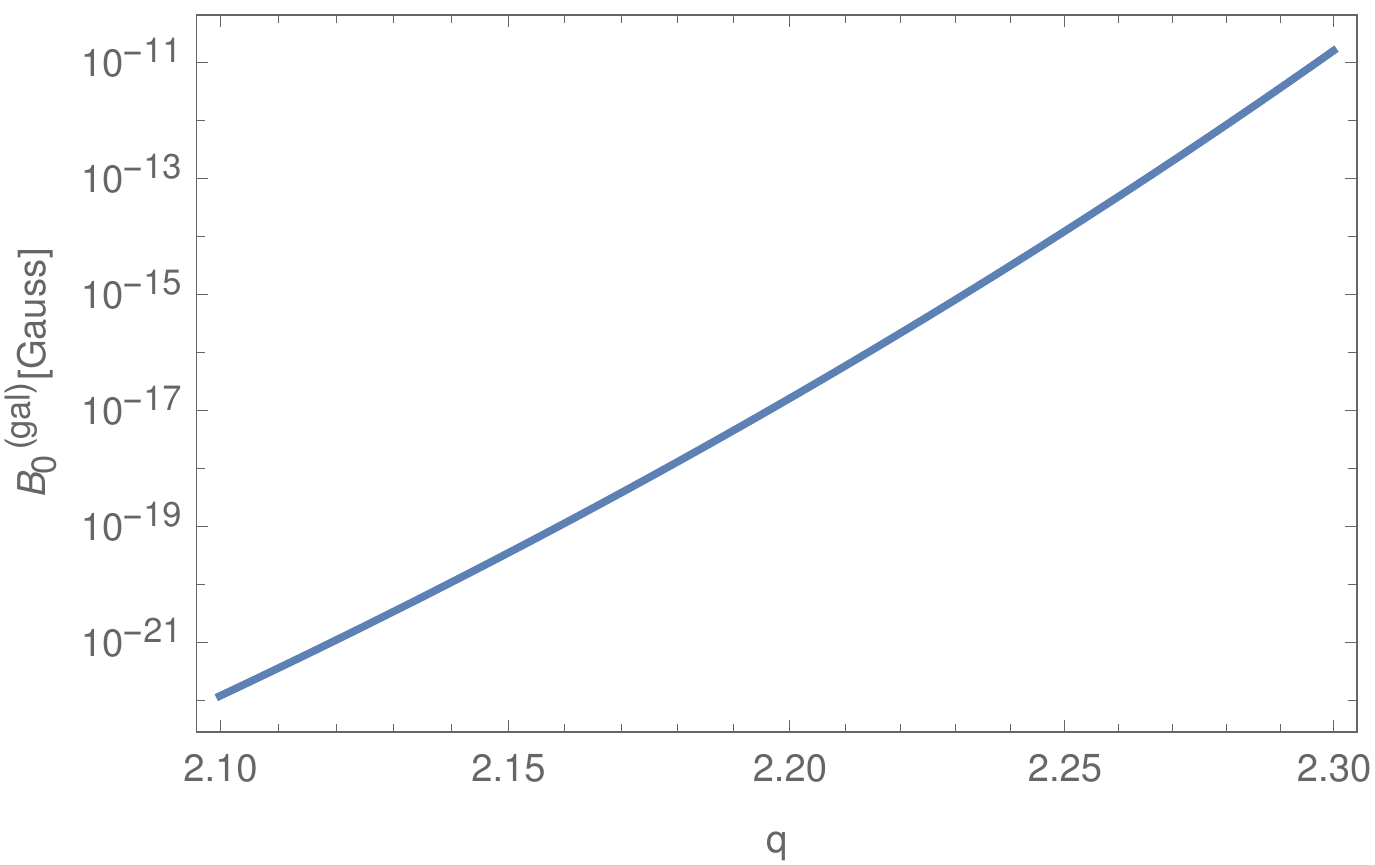}
 \includegraphics[width=3.5in,height=2.5in]{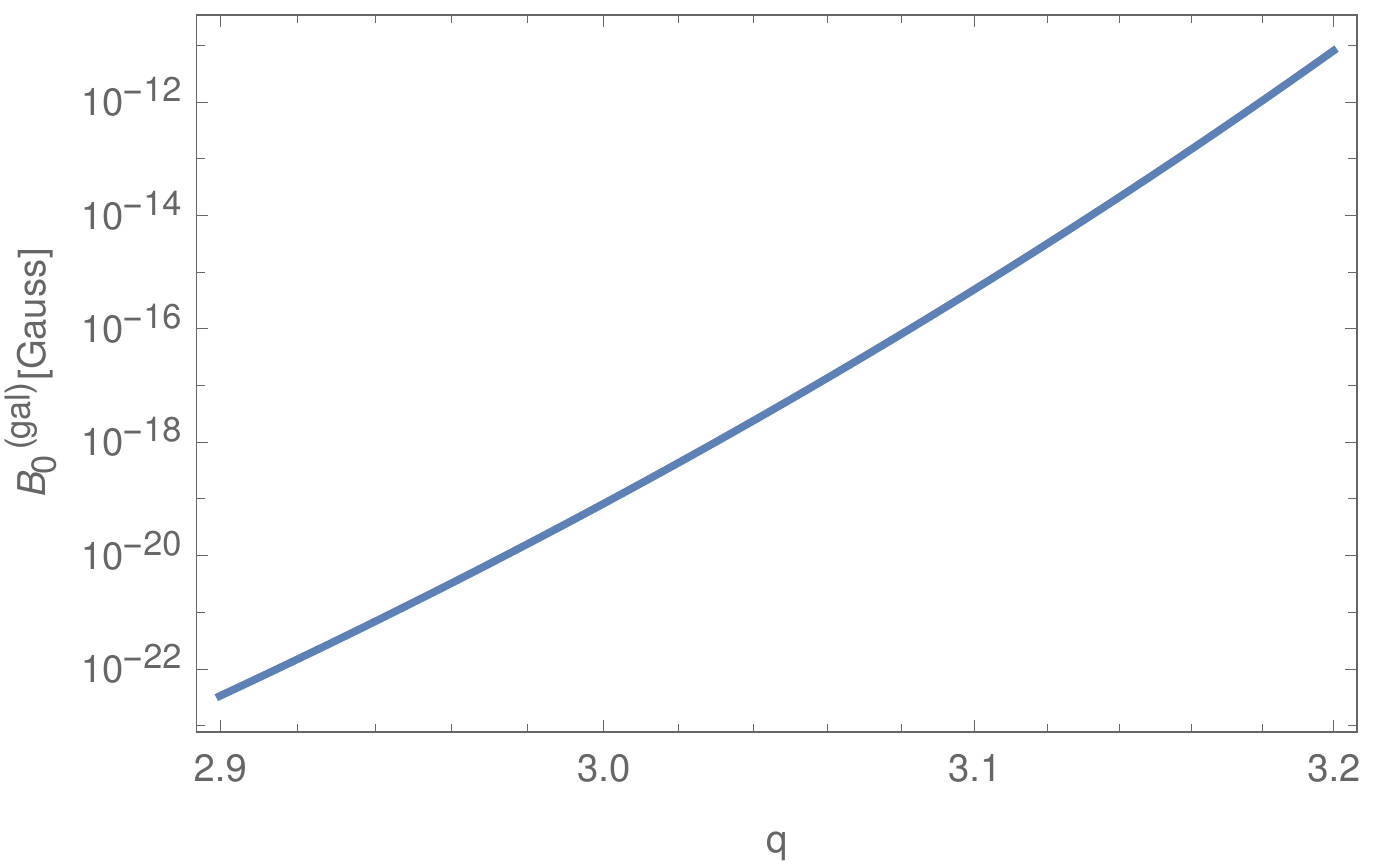}
 \caption{$B_0^\mathrm{(gal)}$ vs $q$. The left and right plot correspond to the case when the EM field couples with the Ricci scalar alone and with the 
 GB scalar alone respectively. In particular, $f(R,\mathcal{G}) = \kappa^{2q}R^q$ for the left plot, and $f(R,\mathcal{G}) = \kappa^{2q}\mathcal{G}^{q/2}$ 
 for the right plot.}
 \label{plot-comparison}
\end{center}
\end{figure}

Before moving to the next section, we would like to explore the following two cases in regard to the non-minimal coupling of the EM field -- 
(1) when the EM field couples with the background Ricci scalar alone, i.e $f(R,\mathcal{G}) = \kappa^{2q}R^q$ and, (2) when the EM field couples with the 
background Gauss-Bonnet scalar alone, i.e $f(R,\mathcal{G}) = \kappa^{2q}\mathcal{G}^{q/2}$. In the first case, by using $R = R(\eta)$ from 
Eq.(\ref{R and G}), we get the non-minimal coupling function (in terms of conformal time) as,
\begin{eqnarray}
 f(R,\mathcal{G}) = \kappa^{2q}\bigg\{\frac{\big[6\beta(\beta+1)\big]^q}{\eta_0^{2q}}\bigg\}\bigg(\frac{-\eta}{\eta_0}\bigg)^{2\epsilon q}~~,
 \label{form of f new1}
\end{eqnarray}
while for the second case, it comes as
\begin{eqnarray}
 f(R,\mathcal{G}) = \kappa^{2q}\bigg\{\frac{\big[-24(\beta+1)^3\big]^{q/2}}{\eta_0^{2q}}\bigg\}\bigg(\frac{-\eta}{\eta_0}\bigg)^{2\epsilon q}~~.
 \label{form of f new2}
\end{eqnarray}
Owing to the above non-minimal coupling functions, the present magnetic field at the galactic scale comes with the same expression as 
of Eq.(\ref{galactic magnetic field}), with $\zeta^2$ is replaced by
\begin{eqnarray}
 \zeta^2 = \left(16\epsilon q\lambda\eta_0\right)\left(\frac{\kappa}{\eta_0}\right)^{2q}\bigg\{\big[6\beta(\beta+1)\big]^q\bigg\}\nonumber
\end{eqnarray}
for the first case, and
\begin{eqnarray}
 \zeta^2 = \left(16\epsilon q\lambda\eta_0\right)\left(\frac{\kappa}{\eta_0}\right)^{2q}\bigg\{\big[-24(\beta+1)^3\big]^{q/2}\bigg\}\nonumber
\end{eqnarray}
for the second case. Therefore we estimate the current magnetic strength in terms of the parameter $q$ for the two 
aforementioned cases. These are depicted in the left and the right plot of Fig.[\ref{plot-comparison}] respectively. 

The Fig.[\ref{plot-comparison}] clearly demonstrates that for both the cases, whether the EM field couples with the Ricci scalar alone or 
with the GB scalar alone, the present magnetic strength comes within the observational constraints for a suitable range of $q$. However it seems 
that the viable range of $q$ changes for the two cases -- in particular, $q$ should lie within $2.1 \leq q \leq 3.0$ for the case when 
$f(R,\mathcal{G}) = \kappa^{2q}R^q$, while for $f(R,\mathcal{G}) = \kappa^{2q}\mathcal{G}^{q/2}$ we get $2.9 \leq q \leq 3.2$ in order to make the 
corresponding magnetogenesis model compatible with the observation. Thus as a whole, the viable range of $q$ for different cases are given by,
\begin{itemize}
 \item $2.1 \leq q \leq 2.26$ for $f(R,\mathcal{G}) = \kappa^{2q}\left(R^q + \mathcal{G}^{q/2}\right)$,\\
 
 \item $2.1 \leq q \leq 2.30$ for $f(R,\mathcal{G}) = \kappa^{2q}R^q$,\\
 
 \item $2.9 \leq q \leq 3.2$ for $f(R,\mathcal{G}) = \kappa^{2q}\mathcal{G}^{q/2}$.
\end{itemize}

It may be observed that the viable range of $q$ does not change considerably for the above cases. This is expected from the fact that $R^{q}$ and 
$\mathcal{G}^{q/2}$ have same order of magnitude during the inflationary era, in particular, $R^{q} \sim \mathcal{G}^{q/2} \sim H_0^{2q}$ during 
the inflation (where $H_0$ is the inflationary Hubble parameter). Therefore one important point to be argued that in the realm 
of higher curvature magnetogenesis scenario, any higher curvature coupling of the EM field, 
that has same order of magnitude with $R^{q}$ (or $\mathcal{G}^{q/2}$) and couples with $F_{\mu\nu}\widetilde{F}^{\mu\nu}$, 
can predict sufficient strength of magnetic field for suitable range of the corresponding model parameter.

\section{Present magnetic strength and magnetic helicity for a Kamionkowski like reheating model with non-zero e-folding number}\label{sec_kamionkowski}
Instead of instantaneous reheating (that we have considered in Sec.[\ref{sec_instantaneous reheating}]), it would be more physical 
if the universe undergoes through a reheating phase which has a non-zero e-fold number. In the case of instantaneous reheating, the 
electrical conductivity immediately turns on after the end of inflation and thus the electric field practically goes to zero. However if the 
reheating phase is considered to have a non-zero e-fold number, then there is no reason to consider a large conductivity immediately after the inflation. 
Indeed, the conductivity remains non-zero and small, due to which, the existence of electric field may induce the magnetic field 
during the stage between the end of inflation and the end of reheating \cite{Kobayashi:2019uqs}. As a result, the magnetic field's strength and 
consequently the helicity spectrum at present universe may become larger compared to those in 
the instantaneous reheating case. In such situation, it is important 
to calculate the magnetic field's strength and the helicity spectrum at current time by considering the reheating phase with a non-zero e-fold number. 
This is our aim in the present section. 

In regard to the reheating dynamics, we consider the conventional reheating mechanism proposed by Kamionkowski et al. \cite{Dai:2014jja}, 
where the main idea is to parametrize the reheating dynamics by -- a non-zero e-fold number, a constant equation of state (EoS) parameter and 
a given reheating temperature. Clearly, the Hubble parameter during the reheating era is connected to that of at the end of inflation by the constant 
EoS parameter (symbolized by $\omega_\mathrm{eff}$), in particular, $H \propto a^{-\frac{3}{2}\left(1+\omega_\mathrm{eff}\right)}$ in 
the reheating stage. Moreover the inflaton energy density is supposed to instantaneously convert to radiation energy density 
after the end of reheating, which indicates the beginning of radiation dominated epoch. 
The characterized quantities of the reheating phase, i.e the e-fold number ($N_\mathrm{re}$) and the 
temperature ($T_\mathrm{re}$) can be expressed in terms of $\omega_\mathrm{eff}$ and of some inflationary parameters by following 
\cite{Dai:2014jja,Cook:2015vqa}, 
\begin{eqnarray}
 N_\mathrm{re} = \frac{4}{\big(1 - 3\omega_\mathrm{eff}\big)}
 \bigg[-\frac{1}{4}\ln{\bigg(\frac{45}{\pi^2g_{re}}\bigg)} - \frac{1}{3}\ln{\bigg(\frac{11g_{s,re}}{43}\bigg)} 
 - \ln{\bigg(\frac{k}{a_0T_0}\bigg)} - \ln{\bigg(\frac{(3H_f^2M_\mathrm{Pl}^2)^{1/4}}{H_0}\bigg)} - N_{f}\bigg]~~~,
 \label{reheating e-folding}
\end{eqnarray}
\begin{eqnarray}
 T_\mathrm{re} = H_0\bigg(\frac{43}{11g_{s,re}}\bigg)^{\frac{1}{3}}\bigg(\frac{a_0T_0}{k}\bigg)\exp{\big[-\big(N_f + N_\mathrm{re}\big)\big]}~~,
 \label{reheating temperature}
\end{eqnarray}
where $T_0 = 2.73\mathrm{K}$ is the present temperature of the universe, 
$\frac{k}{a_0}$ is the pivot scale $\approx 0.05\mathrm{Mpc}^{-1}$ 
in case of large scale modes and $a_0$ is the 
present cosmological scale factor. Here, for simplicity, we have taken both the values of the degrees 
of freedom (d.o.f) as $g_{s,re} = g_{re} \approx 100$, where $g_{s,re}$ is the d.o.f for entropy at reheating and $g_{re}$ is 
the effective number of relativistic species upon 
thermalization. With this reheating model in hand, 
we now solve the electromagnetic mode function and consequently determine the power spectra during the reheating epoch, in the next subsection. 

\subsection{Solution of the mode function and power spectra during the reheating epoch}
We mentioned earlier that the function $f(R,\mathcal{G})$ is considered to be zero after the end of inflation. As a consequence, 
the production of the gauge field from the quantum vacuum stops, which is accounted by the fact that the Bogoliubov coefficients in the post-inflation 
become explicitly time independent and get equal to that of at the end of inflation. Therefore, in the reheating epoch, the EM field follows the 
standard Maxwell's equation, in particular,
\begin{eqnarray}
 A''^{(re)}_{\pm}(k,\eta) + k^2A^{(re)}_{\pm}(k,\eta) = 0~~,
 \label{reheating eom}
\end{eqnarray}
where $A^{(re)}_{\pm}(k,\eta)$ is the EM mode function during the reheating phase. Eq.(\ref{reheating eom}) points that 
the electrical conductivity of the universe during the reheating era is considered to be negligible, in particular to be zero. 
Solving Eq.(\ref{reheating eom}), one gets,
\begin{eqnarray}
 A^{(re)}(k,\eta) = \frac{1}{\sqrt{2k}}\bigg[c_{\pm}~e^{-ik(\eta - \eta_f)} + d_{\pm}~e^{ik(\eta - \eta_f)}\bigg]~~,
 \label{reheating solution1}
\end{eqnarray}
where $c_{\pm}$ and $d_{\pm}$ are integration constants, that can be determined from the continuity conditions of the EM field at the junction between 
the end of inflation and the beginning of reheating, i.e,
\begin{eqnarray}
 A^{(re)}_{\pm}(k,\eta_f) = A_{\pm}(k,\eta_f)~~~~~~~~~~~\mathrm{and}~~~~~~~~~~~A'^{(re)}_{\pm}(k,\eta_f) = A'_{\pm}(k,\eta_f),
 \label{reheating continuity1}
\end{eqnarray}
with $A_{\pm}(k,\eta_f)$ are the EM mode functions at the end of inflation and have the forms as in Eq.(\ref{superhorizon form 1}), see the 
Appendix-A in Sec.[\ref{sec-app1}]. Such continuity relations immediately lead to the integration constants 
in terms of $A_{\pm}(k,\eta_f)$ and $A'_{\pm}(k,\eta_f)$ as,
\begin{eqnarray}
 c_{\pm}&=&\sqrt{\frac{k}{2}}~A_{\pm}(k,\eta_f) + \frac{i}{\sqrt{2k}}~A'_{\pm}(k,\eta_f)\nonumber\\
 d_{\pm}&=&\sqrt{\frac{k}{2}}~A_{\pm}(k,\eta_f) - \frac{i}{\sqrt{2k}}~A'_{\pm}(k,\eta_f)~~.
 \label{continuity2}
\end{eqnarray}
One of the important quantities in magnetogenesis scenario are the 
Bogoliubov coefficients which actually account the production of gauge particles from the quantum vacuum. In particular, due to the interaction of the 
EM field with the background FRW spacetime, the Hamiltonian of the EM field becomes explicit time dependent, which in turn makes the electromagnetic 
vacuum state explicit time dependent. Thus if the EM field starts from the vacuum state at distant past (namely the Bunch-Davies state 
which we consider in the present context), then it will not remain in vacuum at a later time. This corresponds to the production of gauge particles 
from the Bunch-Davies state and the Bogoliubov coefficients provide the number of produced particles at a certain time. More precisely, the Bogoliubov 
coefficients at time $\eta$ relate the mode functions correspond to the vacuum state of distant past with the set of mode functions that lead to the 
instantaneous vacuum of the field at the $\eta$. Based on this, the Bogoliubov coefficients during the reheating epoch, in the present context, 
are determined as,
\begin{eqnarray}
 \alpha_{\pm}(k,\eta)&=&\sqrt{\frac{k}{2}}~A^{(re)}_{\pm}(k,\eta) + \frac{i}{\sqrt{2k}}~A'^{(re)}_{\pm}(k,\eta),\nonumber\\
 \beta_{\pm}(k,\eta)&=&\sqrt{\frac{k}{2}}~A^{(re)}_{\pm}(k,\eta) - \frac{i}{\sqrt{2k}}~A'^{(re)}_{\pm}(k,\eta)~~~~.
 \label{reheating Bogoliubov1}
\end{eqnarray}
Eqs.(\ref{reheating solution1}) and (\ref{reheating Bogoliubov1}) immediately result $\alpha_{\pm}$ and $\beta_{\pm}$ as,
\begin{eqnarray}
 \alpha_{\pm}(k,\eta) = c_{\pm}~e^{-ik(\eta - \eta_f)}~~~~~~~~~~~~,~~~~~~~~~~~~\beta_{\pm}(k,\eta) = d_{\pm}~e^{ik(\eta - \eta_f)},
 \label{reheating Bogoliubov2}
\end{eqnarray}
which relate the Bogoliubov coefficients with the integration constants that appear in the solution of $A^{(re)}_{\pm}(k,\eta)$. 
The absolute value of $\beta_{\pm}$ (i.e $\left|\beta_{\pm}\right|$) during the reheating era seems to be explicit time 
independent and equal to that of at the end of inflation, i.e $\left|\beta_{\pm}(k,\eta)\right| = \left|\beta_{\pm}(k,\eta_f)\right| 
= d_{\pm}$ (where $\eta$ is any time during reheating), from Eq.(\ref{reheating Bogoliubov2}). 
This is a direct consequence of the fact that the coupling function $f(R,\mathcal{G})$ is considered 
to be zero in the reheating stage, due to which the EM action restores the conformal symmetry and thus the EM field equations in the background 
FRW spacetime become similar to that of in the Minkowski spacetime. As a result, the gauge production ceases to occur after the inflation. 
Moreover Eq.(\ref{reheating Bogoliubov2}) leads 
to $c_{\pm} = \alpha_{\pm}(k,\eta_f)$ and $d_{\pm} = \beta_{\pm}(k,\eta_f)$, and thus from Eq.(\ref{reheating solution1}),
\begin{eqnarray}
 A^{(re)}_{\pm}(k,\eta) = \frac{1}{\sqrt{2k}}\bigg[\alpha_{\pm}(k,\eta_f)~e^{-ik(\eta - \eta_f)} + \beta_{\pm}(k,\eta_f)~e^{ik(\eta - \eta_f)}\bigg]~~~.
 \label{reheating solution2}
\end{eqnarray}
Plugging the above solution of $A^{(re)}_{\pm}(k,\eta)$ into Eq.(\ref{power spectra}) along with the relation 
$\left|\alpha_{\pm}\right|^2 - \left|\beta_{\pm}\right|^2 = 1$ yield the magnetic and electric power spectra during the reheating phase as,
\begin{eqnarray}
 \mathcal{P}(\vec{B})&=&\frac{1}{2\pi^2}~\sum_{r=+,-}~\frac{k^5}{a^4}\big|A^{(re)}_{r}(k,\eta)\big|^2\nonumber\\
 &=&\frac{1}{2\pi^2}\sum_{r=+,-}\bigg(\frac{k^4}{a^4}\bigg)\bigg[\big|\alpha_{r}(k,\eta_f)\big|^2 + \big|\beta_r(k,\eta_f)\big|^2 
 + 2\big|\alpha_r(k,\eta_f)~\beta_r(k,\eta_f)\big|~\cos{\big\{\theta^{(r)}_1 - \theta^{(r)}_2 - 2k(\eta - \eta_f)\big\}}\bigg]
 \label{reheating magnetic power spectrum1}
\end{eqnarray}
and
\begin{eqnarray}
 \mathcal{P}(\vec{E})&=&\frac{1}{2\pi^2}~\sum_{r=1,2}~\frac{k^2}{a^4}\big|A_r'^{(re)}(k,\eta)\big|^2\nonumber\\
 &=&\frac{1}{2\pi^2}\sum_{r=+,-}\bigg(\frac{k^4}{a^4}\bigg)\bigg[\big|\alpha_{r}(k,\eta_f)\big|^2 + \big|\beta_r(k,\eta_f)\big|^2 
 - 2\big|\alpha_r(k,\eta_f)~\beta_r(k,\eta_f)\big|~\cos{\big\{\theta^{(r)}_1 - \theta^{(r)}_2 - 2k(\eta - \eta_f)\big\}}\bigg]
 \label{reheating electric power spectrum1}
\end{eqnarray}
respectively, where $\theta^{(r)}_1 = \mathrm{Arg}\left[\alpha_r(k,\eta_f)\right]$ and 
$\theta^{(r)}_2 = \mathrm{Arg}\left[\beta_r(k,\eta_f)\right]$. Consequently, the total electromagnetic power spectrum turns out to be,
\begin{eqnarray}
 \mathcal{P}_{em} = \mathcal{P}(\vec{B}) + \mathcal{P}(\vec{E}) 
 = \frac{1}{\pi^2}~\sum_{r=+,-}~\bigg(\frac{k^4}{a^4}\bigg)\bigg[\big|\alpha_{r}(k,\eta_f)\big|^2 + \big|\beta_r(k,\eta_f)\big|^2\bigg]~~.
 \label{reheating em power spectrum}
\end{eqnarray}
It may be observed from the above expressions that the total EM power spectrum redshifts by $\frac{1}{a^4}$ with the expansion of the universe, 
however the individual magnetic and electric spectra do not evolve as $\frac{1}{a^4}$ due to the presence of the 
time dependent factor $k\left(\eta - \eta_f\right)$ within the cosine argument of Eqs.(\ref{reheating magnetic power spectrum1}) 
and (\ref{reheating electric power spectrum1}) respectively. Thus the comoving EM energy density 
seems to be conserved in the reheating era, which, once again, is a consequence of the fact that the EM action restores the 
conformal symmetry in the post-inflationary epoch. Furthermore Eqs.(\ref{reheating solution2}) and (\ref{helicity power spectrum}) lead to the helicity 
power spectrum in-between the end of inflation and the end of reheating as, 
\begin{eqnarray}
\mathcal{P}_h(k,\eta) &=&\bigg(\frac{k^3}{\pi^2a^3}\bigg)
\bigg\{\big|\beta_+ \big|^2 - \big|\beta_-\big|^2 
+ \sqrt{1 + \big|\beta_+\big|^2}~\big|\beta_+\big|~\cos{\big\{\theta^{(+)}_1 - \theta^{(+)}_2 - 2k(\eta - \eta_f)\big\}}\nonumber\\ 
&-&\sqrt{1 + \big|\beta_-\big|^2}~\big|\beta_-\big|~\cos{\big\{\theta^{(-)}_1 - \theta^{(-)}_2 - 2k(\eta - \eta_f)\big\}}\bigg\}~~.
\label{reheating helicity power spectrum1}
\end{eqnarray}
Here we would like to mention that both the EM mode functions follow the same equation of motion (see Eq.(\ref{reheating eom})) in the 
reheating era, however during inflation, the conformal coupling $f(R,\mathcal{G})$ differs the amplitude of $A_{+}(k,\eta)$ compared to that of 
$A_{-}(k,\eta)$, and as a result, $A_{+}(k,\eta_f) \neq A_{-}(k,\eta_f)$ and $A'_{+}(k,\eta_f) \neq A'_{-}(k,\eta_f)$, i.e positive and negative helicity 
modes obey different initial conditions at the beginning of reheating (recall, $\eta_f$ represents the end instance of inflation or equivalently the 
start of reheating). Thus as a whole, $A_{\pm}(k,\eta)$ have same equations 
of motion but different initial conditions during the reheating stage, in effect of which, 
$A_{+}(k,\eta)$ and $A_{-}(k,\eta)$ get different and makes a non-zero helicity during the same, as given in 
Eq.(\ref{reheating helicity power spectrum1}).

Following, we determine the explicit forms of Bogoliubov coefficients and the factor $k\left(\eta - \eta_f\right)$ present in the expressions 
of various kinds of power spectra.

\begin{itemize}
 \item \textbf{Determination of $\alpha_{\pm}(k,\eta_f)$ and $\beta_{\pm}(k,\eta_f)$}: 
 In view of Eqs.(\ref{continuity2}) and (\ref{reheating Bogoliubov2}), we can express $\alpha_r(k,\eta_f)$ and $\beta_r(k,\eta_f)$ as,
 \begin{eqnarray}
  \alpha_r(k,\eta_f)&=&\sqrt{\frac{k}{2}}\left[A_r(k,\eta_f) + i\frac{dA_r}{d(k\eta)}\bigg|_{\eta_f}\right]~~,\nonumber\\
  \beta_r(k,\eta_f)&=&\sqrt{\frac{k}{2}}\left[A_r(k,\eta_f) - i\frac{dA_r}{d(k\eta)}\bigg|_{\eta_f}\right]~~,
  \label{reheating Bogoliubov 4}
 \end{eqnarray}
with recall, the index $r$ specifies the two polarization modes of the EM field. Using the super-Hubble solution of $A_{\pm}(k,\eta)$, we determine 
the Bogoliubov coefficients as follows:
\begin{eqnarray}
 \alpha_{+}(k,\eta_f)&=&\sqrt{\frac{k}{2}}\left[\left(\frac{C_1 - C_2\cot{\left(\frac{-\pi}{2\alpha}\right)}}{\Gamma\left(1 + \frac{1}{2\alpha}\right)}\right)
 \left(-i\frac{\zeta\sqrt{k}}{2\alpha}\right)^{1/(2\alpha)} + \frac{iH_0}{k}
 \left(\frac{C_2\Gamma\left(\frac{1}{2\alpha}\right)}{\pi}\right)
 \left(-i\frac{\zeta\sqrt{k}}{2\alpha}\right)^{-1/(2\alpha)}\right]~,\nonumber\\
 \beta_{+}(k,\eta_f)&=&\sqrt{\frac{k}{2}}\left[\left(\frac{C_1 - C_2\cot{\left(\frac{-\pi}{2\alpha}\right)}}{\Gamma\left(1 + \frac{1}{2\alpha}\right)}\right)
 \left(-i\frac{\zeta\sqrt{k}}{2\alpha}\right)^{1/(2\alpha)} - \frac{iH_0}{k}
 \left(\frac{C_2\Gamma\left(\frac{1}{2\alpha}\right)}{\pi}\right)
 \left(-i\frac{\zeta\sqrt{k}}{2\alpha}\right)^{-1/(2\alpha)}\right]~,\nonumber\\
 \alpha_{-}(k,\eta_f)&=&\sqrt{\frac{k}{2}}\left[\left(\frac{C_3 - C_4\cot{\left(\frac{-\pi}{2\alpha}\right)}}{\Gamma\left(1 + \frac{1}{2\alpha}\right)}\right)
 \left(\frac{\zeta\sqrt{k}}{2\alpha}\right)^{1/(2\alpha)} + \frac{iH_0}{k}
 \left(\frac{C_4\Gamma\left(\frac{1}{2\alpha}\right)}{\pi}\right)
 \left(\frac{\zeta\sqrt{k}}{2\alpha}\right)^{-1/(2\alpha)}\right]~,\nonumber\\
 \beta_{-}(k,\eta_f)&=&\sqrt{\frac{k}{2}}\left[\left(\frac{C_3 - C_4\cot{\left(\frac{-\pi}{2\alpha}\right)}}{\Gamma\left(1 + \frac{1}{2\alpha}\right)}\right)
 \left(\frac{\zeta\sqrt{k}}{2\alpha}\right)^{1/(2\alpha)} - \frac{iH_0}{k}
 \left(\frac{C_4\Gamma\left(\frac{1}{2\alpha}\right)}{\pi}\right)
 \left(\frac{\zeta\sqrt{k}}{2\alpha}\right)^{-1/(2\alpha)}\right]~.
 \label{reheating Bogoliubov 5}
\end{eqnarray}
The $C_i$ ($i=1,2,3,4$) are given in the Appendix-A. Since $C_1$ and $C_2$ contain complex arguments within the Bessel function, 
one can show from Eq.(\ref{reheating Bogoliubov 5}) that $\left|\beta_{+}(\eta_f)\right|$ is much larger than unity for $q \sim \mathcal{O}(1)$, 
in particular $\left|\beta_{+}(\eta_f)\right| \sim 10^{37}$ for $q = 2.18$ (later we will show that $q=2.18$ leads to sufficient magnetic strength 
at current epoch). On the other hand, $\beta_{-}$ (or $\alpha_{-}$) gets 
suppressed compared to the $\beta_{+}$ (or $\alpha_{+}$), i.e $\left|\beta_{-}\right| \ll \left|\beta_{+}\right|$.

\item \textbf{Determination of ``$\eta - \eta_f$'' during reheating}: The factor $k(\eta - \eta_f)$ in Eqs.(\ref{reheating magnetic power spectrum1}) and 
(\ref{reheating helicity power spectrum1}) leads to the non-conventional dynamics of magnetic and helicity energy density. 
The constant equation of state during reheating dynamics results to this special term as, 
\begin{eqnarray}
 k\big(\eta - \eta_f\big) = \frac{2k}{\big(3\omega_\mathrm{eff} + 1\big)}\bigg[\frac{1}{aH} - \frac{1}{a_fH_f}\bigg],
 \label{reheating conformal 1}
\end{eqnarray}
where we use $\eta - \eta_f = \int^{a}_{a_f}\frac{da}{a^2H}$ and the quantities with suffix 'f' represent the respective quantities at the end of 
inflation.
\end{itemize}

With the above expressions of $\alpha_r(k,\eta_f)$, $\beta_r(k,\eta_f)$ and $k(\eta - \eta_f)$: the magnetic, electric and helicity power spectra 
during the reheating era (from Eqs.(\ref{reheating magnetic power spectrum1}), (\ref{reheating electric power spectrum1}) and 
(\ref{reheating helicity power spectrum1})) turn out to be,
\begin{eqnarray}
 \mathcal{P}(\vec{B}) = \frac{1}{\pi^2}\bigg(\frac{k^4}{a^4}\bigg)\left|\beta_{+}(\eta_f)\right|^2~ 
 \left\{\mathrm{Arg}\left[\alpha_{+}(\eta_f)\beta_{+}^{*}(\eta_f)\right] 
 - \pi - \frac{4k}{3\omega_\mathrm{eff} + 1}\left(\frac{1}{aH} - \frac{1}{a_fH_f}\right)\right\}^2~~,
 \label{reheating magnetic power spectrum2}
\end{eqnarray}
\begin{eqnarray}
 \mathcal{P}(\vec{E}) = \frac{1}{\pi^2}\bigg(\frac{k^4}{a^4}\bigg)\left|\beta_{+}(\eta_f)\right|^2~~,
 \label{reheating electric power spectrum2}
\end{eqnarray}
and
\begin{eqnarray}
 \mathcal{P}_h = \frac{1}{\pi^2}\bigg(\frac{k^3}{a^3}\bigg)\left|\beta_{+}(\eta_f)\right|^2~ 
 \left\{\mathrm{Arg}\left[\alpha_{+}(\eta_f)\beta_{+}^{*}(\eta_f)\right] 
 - \pi - \frac{4k}{3\omega_\mathrm{eff} + 1}\left(\frac{1}{aH} - \frac{1}{a_fH_f}\right)\right\}^2~~,
 \label{reheating helicity power spectrum2}
\end{eqnarray}
respectively, where we neglect the contribution coming from the negative helicity modes. 
Therefore the magnetic power spectrum during the reheating seems to be controlled by two terms: (1) the conventional 
one that redshifts by $1/a^4$ with the expansion of the universe and (2) the term which is proportional to $\left(a^3H\right)^{-2}$ emerged 
from $\left(\frac{k}{aH}\right)^2$ present in the expression of Eq.(\ref{reheating magnetic power spectrum2}). In regard to the helicity power 
spectrum ($\mathcal{P}_h$) during reheating dynamics, Eq.(\ref{reheating helicity power spectrum2}) reveals that, similar to the magnetic spectrum, 
$\mathcal{P}_h$ is also controlled by two terms: (1) the term that evolves by $1/a^3$ and (2) the other one which is proportional to 
$1/\left(a^5H^2\right)$. However due to the constant equation of state, the Hubble parameter during the reheating behaves (in terms of scale factor) 
as $H \propto a^{-\frac{3}{2}\left(1 + \omega_\mathrm{eff}\right)}$ and thus the term $1/\left(a^5H^2\right)$ is further proportional to 
$\propto a^{-2+3\omega_\mathrm{eff}}$. 
The presence of such special terms in the magnetic as well as in the helicity power spectra provide the main difference in the magnetogenesis scenario 
between the instantaneous reheating case and the case where the reheating phase has a non-zero e-fold number. For the magnetic power 
spectrum ($\mathcal{P}(\vec{B})$): in the instantaneous 
reheating case, it evolves by $\frac{1}{a^4}$ from the very end of inflation to the present epoch; however in the case of a non-zero e-fold reheating phase, 
$\mathcal{P}(\vec{B})$ is controlled by Eq.(\ref{reheating magnetic power spectrum2}) 
in-between the end of inflation to the end of reheating and then goes as 
$1/a^4$ from the end of reheating to the present epoch. On other hand, for the helicity power spectrum: in the instantaneous reheating case, the comoving 
helicity remains conserved (i.e $\mathcal{P}_h \propto 1/a^3$) after inflation and until today; however, 
when the reheating phase is considered to have a non-zero e-fold number, $\mathcal{P}_h$ follows 
Eq.(\ref{reheating helicity power spectrum2}) 
from the end of inflation to the end of reheating and then $\mathcal{P}_h \propto 1/a^3$ until the present epoch. 
Finally we would like to mention that unlike to the magnetic or helicity 
power spectra, the electric one decays by $1/a^4$ during the reheating era, as evident from Eq.(\ref{reheating electric power spectrum2}).

\subsection{Current magnetic strength and baryon asymmetry of the universe: Constraints on $\omega_\mathrm{eff}$}
 The presence of a reheating phase with non-zero e-fold number results to the existence of an electric field in the post-inflationary evolution 
 of the universe till the time when the universe becomes purely conductive and this generally occurs at the end of reheating. Such electric field 
 in turn induces the magnetic field according to the Faraday's law of induction \cite{Kobayashi:2019uqs}, 
 due to which, the magnetic field energy density 
 in the reheating epoch redshifts slower than that of in the case of instantaneous reheating. Therefore it is important to investigate whether the 
 presence of a reheating phase having non-zero e-fold yields a sufficient amount of magnetic strength and consequently the helicity at the present time. 
 The Hubble parameter during the reheating phase evolves as 
 $H \propto a^{-\frac{3}{2}\left(1 + \omega_\mathrm{eff}\right)}$, and thus the Hubble parameter at the end of reheating is connected to that of 
 at the end of inflation as, 
\begin{eqnarray}
 H_{re} = H_f\bigg(\frac{a_{re}}{a_f}\bigg)^{-\frac{3}{2}(1 + \omega_{eff})}~~,
 \label{reheating end Hubble parameter}
\end{eqnarray}
where the suffix 're' denotes the end instance of reheating, and 
$\ln{\left(\frac{a_{re}}{a_f}\right)} = N_\mathrm{re}$ represents the e-fold number 
of the reheating era, the $N_\mathrm{re}$ in terms of the reheating parameters is given in Eq.(\ref{reheating e-folding}). The above expression of 
$H_{re}$ immediately leads to the magnetic and helicity power spectra (from Eqs.(\ref{reheating magnetic power spectrum2}) 
and (\ref{reheating helicity power spectrum2})) at the end of reheating as,
\begin{eqnarray}
 \mathcal{P}(\vec{B})\bigg|_{re} = \frac{1}{\pi^2}\left(\frac{k^4}{a_{re}^4}\right)\left|\beta_{+}(\eta_f)\right|^2~ 
 \left\{\mathrm{Arg}\left[\alpha_{+}(\eta_f)\beta_{+}^{*}(\eta_f)\right] 
 - \pi - \frac{4}{3\omega_\mathrm{eff} + 1}
 \left(\frac{k}{a_fH_f}\right)~\exp{\left[\left(\frac{3\omega_\mathrm{eff} + 1}{2}\right)N_\mathrm{re}\right]}\right\}^2~~~.
 \label{reheating magnetic power spectrum3}
\end{eqnarray}
and
\begin{eqnarray}
 \mathcal{P}_h\bigg|_{re} = \frac{1}{\pi^2}\left(\frac{k^3}{a_{re}^3}\right)\left|\beta_{+}(\eta_f)\right|^2~ 
 \left\{\mathrm{Arg}\left[\alpha_{+}(\eta_f)\beta_{+}^{*}(\eta_f)\right] 
 - \pi - \frac{4}{3\omega_\mathrm{eff} + 1}
 \left(\frac{k}{a_fH_f}\right)\exp{\left[\left(\frac{3\omega_\mathrm{eff} + 1}{2}\right)N_\mathrm{re}\right]}\right\}^2~~~.
 \label{reheating helicity power spectrum3}
\end{eqnarray}
respectively, where we use $H_f/H_{re} = \exp{\left[3(1+\omega_\mathrm{eff})N_{re}/2\right]}$ from Eq.(\ref{reheating end Hubble parameter}). 
After the end of reheating, the universe becomes a good conductor, which in turn makes the electric field zero. Moreover due to the 
conformal invariance, the comoving EM energy density as well as the comoving helicity get conserved during the post-reheating era. 
Hence, the present magnetic and the present helicity power spectra is connected to that of at the end of reheating by following:
\begin{eqnarray}
 \mathcal{P}(\vec{B})\bigg|_{0} = \bigg(\frac{a_{re}}{a_0}\bigg)^4~\mathcal{P}(\vec{B})\bigg|_{{re}}
 ~~~~~~~~~~~\mathrm{and}~~~~~~~~~~
 \mathcal{P}_h\bigg|_{0} = \bigg(\frac{a_{re}}{a_0}\bigg)^3~\mathcal{P}_h\bigg|_{{re}}~~,
 \label{reheating magnetic strength 1}
\end{eqnarray}
where the suffix '0' denotes the present time of the universe. Now using Eq.(\ref{reheating magnetic power spectrum3}), 
we get the magnetic field's current amplitude as,
\begin{eqnarray}
 B_0 = \frac{\sqrt{2}}{\pi}\bigg(\frac{k}{a_0}\bigg)^2\left|\beta_{+}(\eta_f)\right|
 \left\{\mathrm{Arg}\left[\alpha_{+}(\eta_f)\beta_{+}^{*}(\eta_f)\right] 
 - \pi - \frac{4}{3\omega_\mathrm{eff} + 1}
 \left(\frac{k}{a_fH_f}\right)\exp{\left[\left(\frac{1+3\omega_\mathrm{eff}}{2}\right)N_\mathrm{re}\right]}\right\}~~,
 \label{reheating magnetic strength 2}
\end{eqnarray}
Furthermore, $\mathcal{P}(\vec{B})\bigg|_{0} = \frac{1}{2}B_0^2$, with 
$B_0$ being the present amplitude of the magnetic field on the scale $k = 0.05\mathrm{Mpc}^{-1}$. 
Considering this generated magnetic fields at the present time play the role 
of seed magnetic fields of galactic magnetic fields and by assuming the magnetic flux conservation \cite{Bamba:2006km}, we can estimate 
the strength of the galactic magnetic fields as,
\begin{eqnarray}
 B_0^\mathrm{(gal)} = \frac{\sqrt{2}}{\pi}\left(\frac{2\pi}{k r_\mathrm{gal}}\right)^2\bigg(\frac{k}{a_0}\bigg)^2\left|\beta_{+}(\eta_f)\right|
 \left\{\mathrm{Arg}\left[\alpha_{+}(\eta_f)\beta_{+}^{*}(\eta_f)\right] 
 - \pi - \frac{4}{3\omega_\mathrm{eff} + 1}
 \left(\frac{k}{a_fH_f}\right)\exp{\left[\left(\frac{1+3\omega_\mathrm{eff}}{2}\right)N_\mathrm{re}\right]}\right\}~.\nonumber\\
 \label{reheating galactic magnetic strength 2}
\end{eqnarray}
Finally, from Eq.(\ref{reheating helicity power spectrum3}), we obtain the helicity power at present epoch as,
\begin{eqnarray}
 \mathcal{P}_h\bigg|_{0} = \frac{1}{\pi^2}\left(\frac{k^3}{a_{0}^3}\right)\left|\beta_+(\eta_f)\right|^2
 \left\{\mathrm{Arg}\left[\alpha_+(\eta_f)\beta_+^{*}(\eta_f)\right] 
 - \pi - \frac{4}{3\omega_\mathrm{eff} + 1}
 \left(\frac{k}{a_fH_f}\right)~\exp{\left[\left(\frac{1+3\omega_\mathrm{eff}}{2}\right)N_\mathrm{re}\right]}\right\}^2~~.
 \label{reheating helicity strength 2}
\end{eqnarray}
Here we need to recall that $\mathcal{P}_h\big|_{0}$ yields the 
net baryon density in present universe ($n_\mathrm{B}$). Therefore it may be observed that both the $B_0^\mathrm{(gal)}$ 
and $n_\mathrm{B}$ depend on the reheating 
dynamics (through $N_\mathrm{re}$ and $\omega_\mathrm{eff}$) as well as on the background inflationary dynamics (through $H_f$ and the absolute value 
of the Bogoliubov coefficients i.e $\left|\beta_{+}(k,\eta_f)\right|^2$). As a result, the magnetic field's current strength and the net baryon density 
of the present universe get a one-to-one correspondence with the reheating equation of state parameter.\\

\begin{figure}[!h]
\begin{center}
 \centering
 \includegraphics[width=3.5in,height=2.5in]{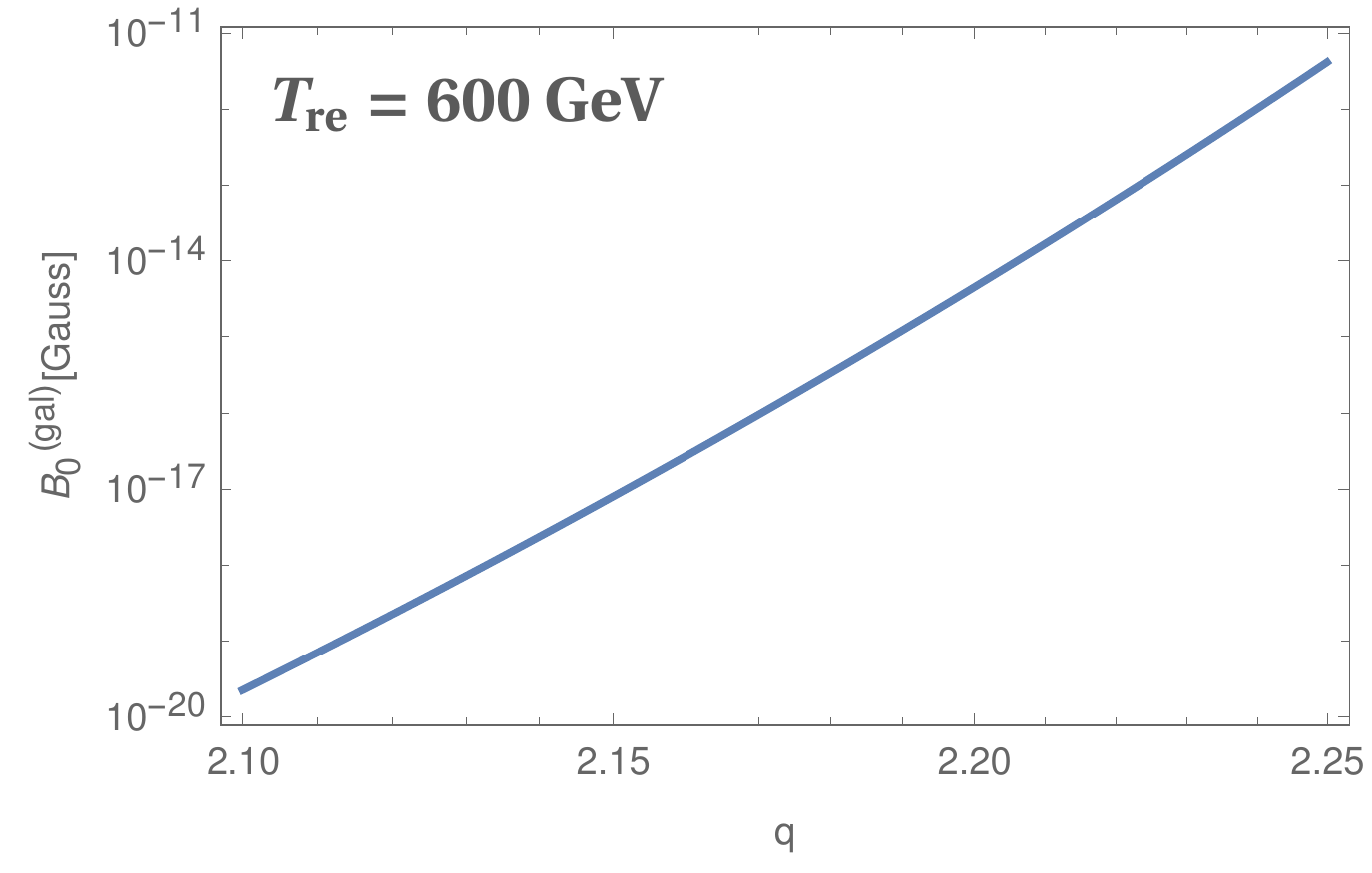}
 \includegraphics[width=3.5in,height=2.5in]{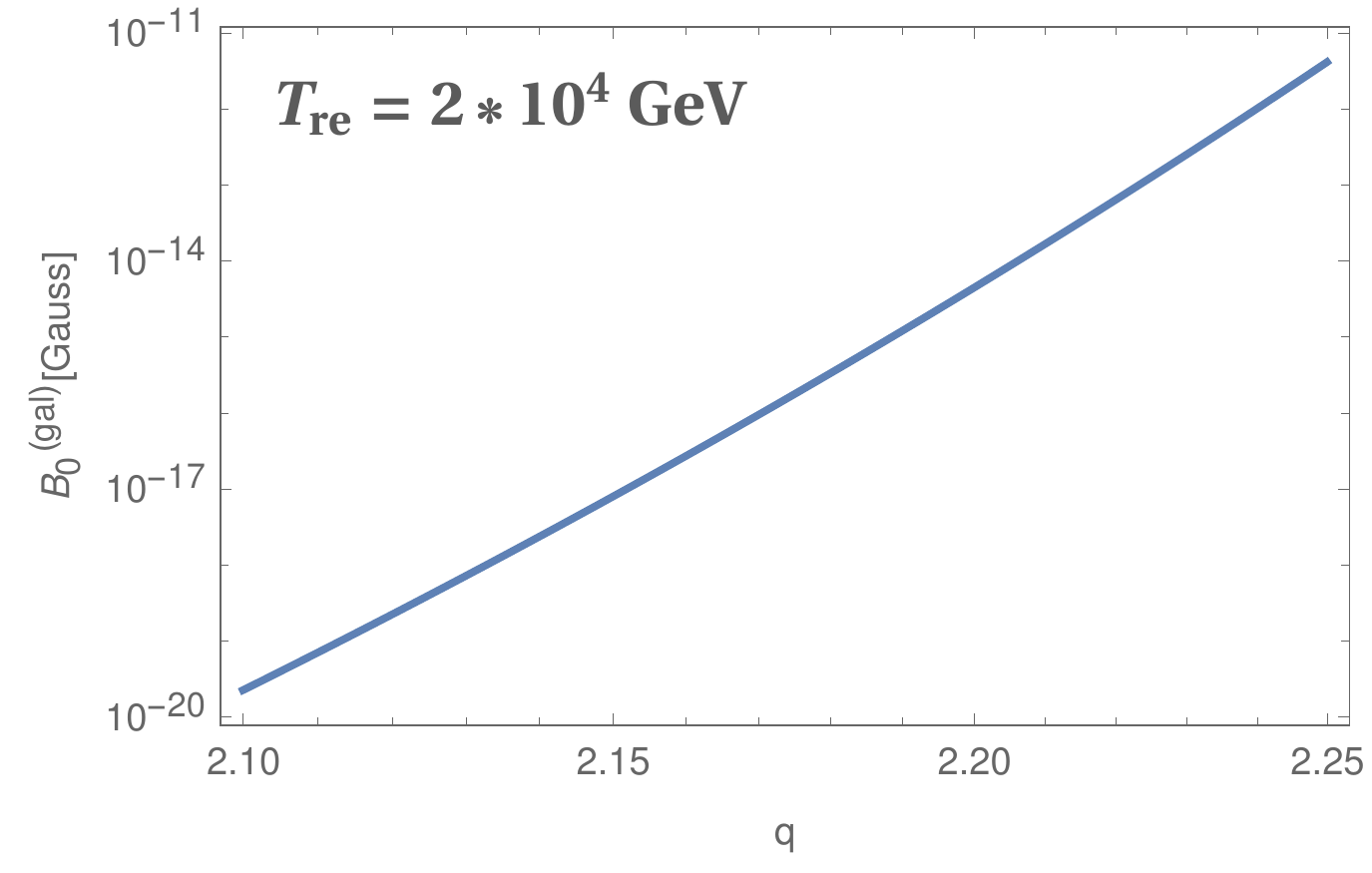}
 \caption{$B_0^\mathrm{(gal)}$ vs $q$ where we take $\lambda = 1$. 
 The left and right plot correspond to the sets 
 $[T_{re},\omega_\mathrm{eff}] = [600\mathrm{GeV},0.16]$ and $[T_{re},\omega_\mathrm{eff}] = [2\times10^{4}\mathrm{GeV},0.14]$ respectively.} 
 \label{plot2}
\end{center}
\end{figure}

Thus as a whole, Eqs.(\ref{reheating galactic magnetic strength 2}) and (\ref{reheating helicity strength 2}) 
are the key equations to determine the magnetic field's 
current strength and the net baryon density of the universe. Having obtained such expressions, we now confront the model with the observations 
which put a constraint on the present magnetic strength as $10^{-22}\mathrm{G} \lesssim B_0^\mathrm{(gal)} \lesssim 10^{-10}\mathrm{G}$. 
In regard to this, we take $H_0 = 1.6\times10^{13}\mathrm{GeV}$, $\epsilon = 0.1$ and $N_f = 51$, which immediately leads to 
$H_f = 9\times10^{10}\mathrm{GeV}$. Moreover the parameter $q$ is constrained by $q \leq 3.3$ in 
order to resolve the backreaction issue. With such considerations and by using Eq.(\ref{reheating galactic magnetic strength 2}), 
we plot the current magnetic strength ($B_0^\mathrm{(gal)}$) vs $q$ for two different reheating temperatures ($T_{re}$), see Fig.[\ref{plot2}]. 
In particular, we consider $T_{re} = 6\times10^2\mathrm{GeV}$ and $T_{re} = 2\times10^{4}\mathrm{GeV}$, which correspond to 
$\omega_\mathrm{eff} = 0.16$ and $0.14$ respectively. For detailed numerical estimation of $B_0^\mathrm{(gal)}$, see the Appendix-B in 
Sec.[\ref{sec-app2}].\\

The Fig.[\ref{plot2}] clearly demonstrates that 
$B_0^\mathrm{(gal)}$ lies within the observational constraints if the model parameter $q$ satisfies: $2.1 \leq q \leq 2.25$ for both the values 
of $T_{re}$ considered.\\

By comparing Fig.[\ref{plot-present-mag-str}] and Fig.[\ref{plot2}], it is clear that 
the viable range of $q$ remains almost same in both the instantaneous and Kamionkowski reheating scenario, or we may argue that 
the generation of helical magnetic fields in the present context is not much affected by the reheating era. 
This is due to the fact that the present magnetogenesis model 
does not produce sufficient hierarchy between the electric and magnetic fields at the end of inflation, 
and thus the electric field is not able to sufficiently induce (or enhance) the magnetic field during the Kamionkowski reheating stage 
-- which in turn makes both the reheating cases almost similar from the perspective of magnetic field's evolution. The demonstration goes as follows: 
actually the information of the reheating era in the magnetic power spectrum comes through the term 
$(-k\eta_f)e^{\left(\frac{3\omega_\mathrm{eff} + 1}{2}\right)N_\mathrm{re}}$ present 
in Eq.(\ref{reheating galactic magnetic strength 2}). However due to the reason that the electric and magnetic fields do not have much hierarchy 
at the end of inflation (particularly $\mathcal{P}(\vec{E})/\mathcal{P}(\vec{B})\big|_{f} \sim 10^{4}$, where the suffix 'f' represents the end instant of 
inflation), the term $(-k\eta_f)e^{\left(\frac{3\omega_\mathrm{eff} + 1}{2}\right)N_\mathrm{re}}$ 
gets heavily suppressed compared to the other term sitting within the paranthesis of Eq.(\ref{reheating galactic magnetic strength 2}), i.e. 
$\left|\mathrm{Arg}\left[\alpha_{+}(\eta_f)\beta_{+}^{*}(\eta_f)\right] - \pi\right|$, namely
\begin{eqnarray}
 \left(\frac{k}{a_fH_f}\right)~\exp{\left[\left(\frac{1+3\omega_\mathrm{eff}}{2}\right)N_\mathrm{re}\right]} \ll 
 \left|\mathrm{Arg}\left[\alpha_{+}(\eta_f)\beta_{+}^{*}(\eta_f)\right] - \pi\right|~\label{new}
\end{eqnarray}
(see the estimation in the Appendix-B). As a result, the $B_0^\mathrm{(gal)}$ of Eq.(\ref{reheating galactic magnetic strength 2}) 
becomes almost independent of $\omega_\mathrm{eff}$. In particular owing to Eq.(\ref{new}), 
the scaling of the magnetic spectrum during the Kamionkowski reheating stage is approximately given by $\sim a^{-4}$ -- which is 
independent of $\omega_\mathrm{eff}$ and same as the magnetic field scales after the 
end of inflation in the instantaneous reheating case. The above arguments reveal that why the generation of helical magnetic fields 
in the present context is not much affected by the presence of the Kamionkowski reheating stage characterized by $\omega_\mathrm{eff}$. Here we would 
like to mention that this 
is unlike to the higher curvature magnetogenesis scenario proposed in \cite{Bamba:2020qdj} where the spacetime curvature couples with 
$F_{\mu\nu}F^{\mu\nu}$ (i.e the magnetic field is $not~helical$ in nature) and consequently, the electric field at the end of inflation 
becomes much stronger that that of the magnetic field, in particular, the ratio of electric and magnetic field at the end of inflation 
comes as $\mathcal{P}(\vec{E})/\mathcal{P}(\vec{B})\big|_{f} \sim e^{2N_\mathrm{f}} \gg 1$ (with $N_\mathrm{f}$ being the inflationary e-folding number). 
Due to such strong hierarchy between the electric and the magnetic 
fields, the relative phase 
between $\alpha_{+}(\eta_f)$ and $\beta_{+}(\eta_f)$ becomes equal to $\pi$ from Eq.(\ref{reheating Bogoliubov 4}), and consequently 
the term $(-k\eta_f)e^{\left(\frac{3\omega_\mathrm{eff} + 1}{2}\right)N_\mathrm{re}}$ dominates over 
$\left|\mathrm{Arg}\left[\alpha_{+}(\eta_f)\beta_{+}^{*}(\eta_f)\right] - \pi\right|$. In effect, the magnetic spectrum during the Kamionkowski 
reheating stage scales as $\left(a^3H\right)^{-2}e^{\left(3\omega_\mathrm{eff} + 1\right)N_\mathrm{re}}$ -- which indeed depends on the reheating 
equation of state parameter. Thus in the magnetogenesis scenario \cite{Bamba:2020qdj}, the Kamionkowski reheating stage shows considerable effects 
on the magnetic field's evolution, in particular, the magnetic field seems to be enhanced in the Kamionkowski reheating case compared to that of 
in the instantaneous reheating case. The above arguments clearly demonstrate that the hierarchy between electric and magnetic fields at the end of 
inflation plays a significant role whether the Kamionkowski reheating phase affects the generation of magnetic field in 
a magnetogenesis scenario or not. In the present context of higher curvature helical magnetogenesis scenario, the electric and the 
magnetic fields do not get a considerable hierarchy at the end of inflation, and consequently, the generation of the helical magnetic field is not much 
affected by the reheating physics.\\

In regard to the generation of BAU from helical magnetic fields, the net baryonic number density ($n_\mathrm{B}$) can be determined as,
\begin{eqnarray}
 n_\mathrm{B} = 
 \left(\frac{n_\mathrm{f}}{2}\right)\left(\frac{g'^2}{4\pi^4}\right)\left(\frac{k_0^3}{a_{0}^3}\right)\left|\beta_+(\eta_f)\right|^2
 \left\{\mathrm{Arg}\left[\alpha_+(\eta_f)\beta_+^{*}(\eta_f)\right] 
 - \pi - \frac{4}{3\omega_\mathrm{eff} + 1}
 \left(\frac{k_0}{a_fH_f}\right)~\exp{\left[\left(\frac{1+3\omega_\mathrm{eff}}{2}\right)N_\mathrm{re}\right]}\right\}^2~~,
 \nonumber\\
 \label{baryon number-1-reh}
\end{eqnarray}
where, recall that $k_0 = a_0H_0$, i.e $k_0^{-1}$ is the present horizon scale. Owing to Eq.(\ref{reheating galactic magnetic strength 2}), 
the above expression of $n_\mathrm{B}$ can be equivalently written as,
\begin{eqnarray}
 n_\mathrm{B} = \left(\frac{n_\mathrm{f}}{2}\right)\left(\frac{g'^2}{4\pi^2}\right)\left(\frac{a_0}{k_0}\right)
 \left(\frac{k r_\mathrm{gal}}{2\pi}\right)^4\left(B_0^\mathrm{(gal)}\right)^2~~.
 \label{baryon number-2-reh}
\end{eqnarray}
In regard to estimate the net baryon density, we will use Eq.(\ref{baryon number-2-reh}) where $n_\mathrm{B}$ has been expressed in terms of 
the $B_0^\mathrm{(gal)}$. As mentioned earlier, we are interested to determine $n_\mathrm{B}/s_0$ where $s_0$ is given earlier in 
Eq.(\ref{entropy}). With the set of values 
that we considered earlier (i.e $H_0 = 1.6\times10^{13}\mathrm{GeV}$, $\epsilon = 0.1$, and $N_\mathrm{f} = 51$), 
we estimate $n_\mathrm{B}/s_0$ for 
different values of $B_0$, with $T_{re} = 6\times10^{2}\mathrm{GeV}$ and $\frac{k_0}{a_0} = 3.2\times10^{-40}\mathrm{GeV}$. 
Such estimation of $n_\mathrm{B}/s_0$ in Kamionkowski reheating scenario resembles with that of presented in the 
Table[\ref{Table-1}] which clearly demonstrates that the magnetic fields of current strength 
$B_0^\mathrm{(gal)} \sim 10^{-13}\mathrm{G}$ 
leads to the resultant value of $n_\mathrm{B}/s_0$ as of the order $\sim 10^{-10}$. 
 
 Thereby similar to the instantaneous reheating case, the current magnetic strength and the net baryonic number density in the Kamionkowski reheating 
 scenario are found to be compatible with the observational constraints, provided the parameter $q$ lies within $2.1 \leq q \leq 2.5$.

 \section{Conclusion}
We provide a viable helical magnetogenesis scenario from inflation and explain the baryon asymmetry of the universe. 
The electromagnetic (EM) field couples with the background higher 
curvature term(s), in particular with the Ricci scalar and with the Gauss-Bonnet (GB) invariant, via the dual field tensor. Such non-minimal coupling of the EM 
field breaks the conformal and the parity symmetries, however preserves the $U(1)$ invariance, of the electromagnetic action. As a result, the positive 
and negative helicity modes get different amplitudes along their cosmological evolution, 
and thus the magnetic field becomes helical in nature, which is further connected to the net baryonic number density of the universe. 
The background spacetime is governed by the well studied $\alpha$-attractor scalar-tensor theory 
which is known to lead to a viable inflationary scenario consistent with the recent Planck data. The model 
has some good features: first -- during the early universe when the curvature is large, the conformal breaking coupling remains significant and results 
to a non-trivial contribution to the EM field equations, however during the late times, the curvature becomes low, and as a consequence, the conformal 
breaking term does not actually contribute particularly from the end of inflation. 
Second -- due to the fact that the non-minimal coupling of the EM field 
occurs via the dual field tensor, the kinetic term of the EM field remains canonical even in presence of such non-minimal coupling. 
This leads to the resolution 
of strong coupling problem in the present model. As a third strong point of our model -- the EM field is found to have a negligible backreaction 
on the background spacetime and consequently the backreaction issue is resolved in the model.

In such a scenario, we explore the cosmological evolution of electric and magnetic fields, starting from the inflationary era to the present epoch. 
Consequently we address the helicity power spectrum, that arises due to the difference in the amplitudes between the positive and negative 
helicity modes of the EM field. During the evolution, the universe enters to a reheating era after the end of inflation and depending on the reheating 
dynamics, we consider two different cases: (1) the instantaneous reheating case where the universe instantaneously jumps to the radiation dominated epoch 
as the inflation ends, i.e the e-fold number of the instantaneous reheating era is zero. (2) As a second case, the reheating phase 
is considered to have a non-zero e-fold number, in particular, we consider the conventional reheating mechanism proposed by Kamionkowski et al. 
\cite{Dai:2014jja}, where the reheating phase is parametrized by 
a constant equation of state parameter ($\omega_\mathrm{eff}$) and a reheating temperature ($T_{re}$).

In the case of instantaneous reheating, as the universe suddenly jumps from inflation to radiation era, the conductivity of the universe 
becomes huge in the post-inflationary stage and thus the electric field dies out quickly. Moreover, the EM action restores the conformal symmetry after 
inflation and consequently the EM energy density evolves as $a^{-4}$ with the expansion of the universe, 
or equivalently the magnetic energy density redshifts as $a^{-4}$ as the electric field practically vanishes. Consequently the helicity power 
spectrum goes by $a^{-3}$, i.e the comoving helicity remains conserved with cosmic time in the post-inflation phase. As a result, the 
theoretical predictions for the galactic magnetic field's current strength ($B_0^\mathrm{(gal)}$) is found to be consistent with the observational 
constraints ($10^{-22}G < B_0^\mathrm{(gal)} < 10^{-10}G$) for a suitable regime of the parameter $q$, in particular for 
$2.1 \leq q \leq 2.6$. Such range of $q$ also results to a negligible backreaction of the EM field on the background spacetime. Therefore the present 
scenario is able to predict sufficient magnetic strength at present epoch and concomitantly resolves the backreaction issue. Furthermore 
we have found that the magnetic fields with strength $B_0^\mathrm{(gal)} \sim 10^{-13}\mathrm{G}$ at present time can lead to 
the resultant value of the ratio of the baryonic number density to the entropy density, i.e $n_\mathrm{B}/s_0$, 
as large as $10^{-10}$, which is consistent with the observational data of BAU.

Instead of the instantaneous reheating, it would be more physical if the universe
undergoes through a reheating phase which has a non-zero e-fold number. 
As mentioned, we consider a Kamionkowski like reheating mechanism where, due to the constant EoS parameter symbolized by 
$\omega_\mathrm{eff}$, the Hubble parameter during the reheating era is given by $H \propto a^{-\frac{3}{2}\left(1+\omega_\mathrm{eff}\right)}$. 
After the end of reheating, the universe 
enters to a radiation era and the conductivity gets a huge value, which in turn makes the electric field practically zero. 
Specifically, in regard to the magnetic power spectrum -- it evolves 
by Eq.(\ref{reheating magnetic power spectrum2}) during the reheating stage, and after that, 
as $1/a^4$ till the present epoch. On other hand, the helicity power spectrum 
follows Eq.(\ref{reheating helicity power spectrum2}) in the reheating era, and then, as $1/a^3$ from the end of reheating to the present time. 
It turns out that similar to the instantaneous reheating case, the current magnetic strength at the galactic scale ($B_0^\mathrm{(gal)}$) 
and the net baryon density ($n_\mathrm{B}$) in Kamionkowski reheating scenario are compatible with the observational 
constraints. In particular, we get $2.1 \leq q \leq 2.5$, in order 
to make compatible the theoretical predictions of the present model with the observations of $B_0^\mathrm{(gal)}$ and $n_\mathrm{B}/s_0$. 
It is evident that the viable range of $q$ remains almost same in both the instantaneous and Kamionkowski reheating scenario. This is 
due to the fact that the present magnetogenesis model does not produce sufficient hierarchy between the electric and 
magnetic fields at the end of inflation, and thus the electric field is not able to sufficiently induce (or enhance) the 
magnetic field during the Kamionkowski reheating stage. This in turn makes both the reheating cases almost similar from the perspective 
of magnetic field's evolution.

\section{Appendix-A: Solutions and power spectrum of the EM field during inflation}\label{sec-app1}

To solve Eq.(\ref{FT eom region II}), we introduce a dimensionless variable,
\begin{eqnarray}
 \tau = \left(\frac{-\eta_0}{\eta}\right)^{\alpha}~~.
 \label{app1-1}
\end{eqnarray}
In terms of $\tau$, Eq.(\ref{FT eom region II}) turns out to be,
\begin{eqnarray}
 \alpha^2\frac{d^2A_{\pm}}{d\tau^2} + \left(\frac{\alpha(\alpha + 1)}{\tau}\right)\frac{dA_{\pm}}{d\tau} \mp k\zeta^2A_{\pm(k,\tau)} = 0~~.
 \label{app1-2}
\end{eqnarray}
The above equation is a Bessel differential equation and has a complete solution as,
\begin{eqnarray}
A_{+}(k,\eta)&=&\tau^{-1/(2\alpha)}\left\{C_1~
 J_{\frac{1}{2\alpha}}\left(-i\frac{\zeta\sqrt{k}}{\alpha}\tau\right) 
 + C_2~Y_{\frac{1}{2\alpha}}\left(-i\frac{\zeta\sqrt{k}}{\alpha}\tau\right)\right\}~~,\nonumber\\
 A_{-}(k,\eta)&=&\tau^{-1/(2\alpha)}\left\{C_3~
 J_{\frac{1}{2\alpha}}\left(\frac{\zeta\sqrt{k}}{\alpha}\tau\right) 
 + C_4~Y_{\frac{1}{2\alpha}}\left(\frac{\zeta\sqrt{k}}{\alpha}\tau\right)\right\}~~,
 \label{app1-3}
\end{eqnarray}
which, in terms of $\eta$, resembles with the form given in Eq.(\ref{sol1}). By matching $A_{+}(k,\eta)$ and $A_{+}'(k,\eta)$ at 
$\eta=\eta_{*}$ (recall, $\eta_{*}$ is the horizon crossing instant of $k$-th mode), we get $C_i$ ($i=1,2$) as follows:
\begin{eqnarray}
 C_1&=&\frac{-1}{2\alpha\sqrt{-2k\eta_{*}/\eta_0}}\left[i\pi e^{-ik\eta_{*}}\left\{-\zeta\sqrt{k}\tau_{*}~
 Y_{1+\frac{1}{2\alpha}}\left(-i\frac{\zeta\sqrt{k}}{\alpha}\tau_{*}\right) 
 + k\eta_{*}~Y_{\frac{1}{2\alpha}}\left(-i\frac{\zeta\sqrt{k}}{\alpha}\tau_{*}\right)\right\}\right]~~,\nonumber\\
C_2&=&\frac{-1}{4\alpha\sqrt{-2k\eta_{*}/\eta_0}}\bigg[i\pi e^{-ik\eta_{*}}\bigg\{-\zeta\sqrt{k}\tau_{*}
 \left(J_{-1+\frac{1}{2\alpha}}\left(-i\frac{\zeta\sqrt{k}}{\alpha}\tau_{*}\right) 
 - J_{1+\frac{1}{2\alpha}}\left(-i\frac{\zeta\sqrt{k}}{\alpha}\tau_{*}\right)\right)\nonumber\\
 &+&\left(i - 2k\eta_{*}\right)~J_{\frac{1}{2\alpha}}\left(-i\frac{\zeta\sqrt{k}}{\alpha}\tau_{*}\right)
 \bigg\}\bigg]~~.
 \label{app1-4}
 \end{eqnarray}
 Similarly the matching conditions of $A_{-}(k,\eta)$ and $A_{-}'(k,\eta)$ lead to $C_3$, $C_4$ as,
 \begin{eqnarray}
 C_3&=&\frac{-1}{2\alpha\sqrt{-2k\eta_{*}/\eta_0}}\left[-\pi e^{-ik\eta_{*}}\left\{-\zeta\sqrt{k}\tau_{*}~
 Y_{1+\frac{1}{2\alpha}}\left(\frac{\zeta\sqrt{k}}{\alpha}\tau_{*}\right) 
 - ik\eta_{*}~Y_{\frac{1}{2\alpha}}\left(\frac{\zeta\sqrt{k}}{\alpha}\tau_{*}\right)\right\}\right]~~,\nonumber\\
 C_4&=&\frac{-1}{2\alpha\sqrt{-2k\eta_{*}/\eta_0}}\left[\pi e^{-ik\eta_{*}}\left\{-\zeta\sqrt{k}\tau_{*}~
 J_{1+\frac{1}{2\alpha}}\left(\frac{\zeta\sqrt{k}}{\alpha}\tau_{*}\right) 
 - ik\eta_{*}~J_{\frac{1}{2\alpha}}\left(\frac{\zeta\sqrt{k}}{\alpha}\tau_{*}\right)\right\}\right]~.
 \label{app1-5}
\end{eqnarray}
In the above expressions, $\tau_{*} = \left(-\eta_0/\eta_{*}\right)^{\alpha}$ and 
$k\eta_{*} = -(1+\epsilon)$, as shown in Eq.(\ref{horizon crossing condition}). 
From Eq.(\ref{sol1}), 
$A_{\pm}(k,\eta)$ in the superhorizon limit (i.e $|k\eta| \ll 1$) come as,
\begin{eqnarray}
 \lim_{|k\eta| \ll 1} A_+(k,\eta)&=&\left(\frac{C_1 - C_2\cot{\left(\frac{-\pi}{2\alpha}\right)}}{\Gamma\left(1 + \frac{1}{2\alpha}\right)}\right)
 \left(-i\frac{\zeta\sqrt{k}}{2\alpha}\right)^{1/(2\alpha)}~~,\nonumber\\
 \lim_{|k\eta| \ll 1} A_-(k,\eta)&=&\left(\frac{C_3 - C_4\cot{\left(\frac{-\pi}{2\alpha}\right)}}{\Gamma\left(1 + \frac{1}{2\alpha}\right)}\right)
 \left(\frac{\zeta\sqrt{k}}{2\alpha}\right)^{1/(2\alpha)}~~.
 \label{superhorizon form 1}
 \end{eqnarray}
 Moreover $A_{\pm}'(k,\eta)$ (which are necessary for electric power spectrum) in the superhorizon regime are obtained as, 
 \begin{eqnarray}
 \lim_{|k\eta| \ll 1} \frac{dA_+}{d(k\eta)}&=&H_0\left(\frac{C_2\Gamma\left(\frac{1}{2\alpha}\right)}{\pi}\right)
 \left(-i\frac{\zeta\sqrt{k}}{2\alpha}\right)^{-1/(2\alpha)}~~,\nonumber\\
 \lim_{|k\eta| \ll 1} \frac{dA_-}{d(-k\eta)}&=&H_0\left(\frac{C_4\Gamma\left(\frac{1}{2\alpha}\right)}{\pi}\right)
 \left(\frac{\zeta\sqrt{k}}{2\alpha}\right)^{-1/(2\alpha)}~~.\nonumber\\
 \label{superhorizon form 2}
\end{eqnarray}
Using such expressions, one arrives the electric and magnetic power spectra during inflation, as obtained in 
Eq.(\ref{electric power spectrum 1}) and Eq.(\ref{magnetic power spectrum 1}) respectively.

\section{Appendix-B: Numerical estimation of magnetic strength}\label{sec-app2}

From Eq.(\ref{reheating galactic magnetic strength 2}), 
the magnetic field strength at present epoch (in presence of Kamionkowski reheating phase) is given by,
\begin{eqnarray}
 B_0^\mathrm{(gal)} = \frac{\sqrt{2}}{\pi}\left(\frac{2\pi}{k r_\mathrm{gal}}\right)^2\bigg(\frac{k}{a_0}\bigg)^2\left|\beta_{+}(\eta_f)\right|
 \left\{\mathrm{Arg}\left[\alpha_{+}(\eta_f)\beta_{+}^{*}(\eta_f)\right] 
 - \pi - \frac{4}{3\omega_\mathrm{eff} + 1}
 \left(\frac{k}{a_fH_f}\right)\exp{\left[\left(\frac{1+3\omega_\mathrm{eff}}{2}\right)N_\mathrm{re}\right]}\right\}~~,
 \label{reheating galactic magnetic strength 2-app}
\end{eqnarray}
where $k/(a_fH_f) = -k\eta_f$. Moreover 
$N_{re}$ represents the e-fold number of the reheating phase, and given in Eq.(\ref{reheating e-folding}) as,
\begin{eqnarray}
 N_\mathrm{re} = \frac{4}{\big(1 - 3\omega_\mathrm{eff}\big)}
 \bigg[-\frac{1}{4}\ln{\bigg(\frac{45}{\pi^2g_{re}}\bigg)} - \frac{1}{3}\ln{\bigg(\frac{11g_{s,re}}{43}\bigg)} 
 - \ln{\bigg(\frac{k}{a_0T_0}\bigg)} - \ln{\bigg(\frac{(3H_f^2M_\mathrm{Pl}^2)^{1/4}}{H_0}\bigg)} - N_{f}\bigg]~.
 \label{reheating e-folding-app}
\end{eqnarray}
Now the estimations for different quantities are,\\

\begin{table}[h]
  \centering
 {%
  \begin{tabular}{|c|c|}
   \hline 
    Quantities & Corresponding value (unit)\\
   \hline
   $[q,~\lambda]$ & $[2.18,~1]$ \\
   \hline
   $k/a_0$ & $0.05\mathrm{Mpc}^{-1} = 3.2\times10^{-40}\mathrm{GeV}$ \\
   \hline
   $[H_0,~\epsilon,~N_f]$ & $[1.6\times10^{13}\mathrm{GeV},~0.1,~51]$ \\
   \hline
   $[T_{re},~\omega_\mathrm{eff}]$ & $[6\times10^{2}\mathrm{GeV},~0.16]$ \\
   \hline
   $T_0$ & $2.73\mathrm{K} = 2.35\times10^{-13}\mathrm{GeV}$ \\
   \hline
   \hline
  \end{tabular}%
 }
  \caption{Various quantities and their corresponding values with appropriate unit.}
  \label{Table-2}
 \end{table}
 
where we use the conversion $1\mathrm{K} = 8.6\times10^{-14}\mathrm{GeV}$. Such considered values immediately lead to $N_\mathrm{re} = 33$ 
and $H_{re} = 6\times10^{-13}\mathrm{GeV}$ respectively. Consequently, we estimate the following quantities:
\begin{eqnarray}
 \left|\beta_{+}(\eta_f)\right|&=&1.8\times10^{37}~~,\nonumber\\
 \left|\mathrm{Arg}\left[\alpha_{+}(\eta_f)\beta_{+}^{*}(\eta_f)\right] - \pi\right|&=&3\times10^{-3}~~,\nonumber\\
 \left(\frac{4k}{a_fH_f}\right)\exp{\left[\left(\frac{1+3\omega_\mathrm{eff}}{2}\right)N_\mathrm{re}\right]}&=&1.3\times10^{-9}~~.
 \label{app2-1}
\end{eqnarray}

Plugging all such values in Eq.(\ref{reheating galactic magnetic strength 2-app}), one gets,
\begin{eqnarray}
 B_0^\mathrm{(gal)}&\sim&10^{10}\times10^{-80}\times10^{37}\times10^{-3}\times\left(\frac{1}{1.95\times10^{-20}}\right)\mathrm{Gauss}~~,\nonumber\\
 &\sim&10^{-16}\mathrm{Gauss}~~.
 \label{reheating galactic magnetic strength 3-app}
\end{eqnarray}
Similarly, one can estimate $B_0^\mathrm{(gal)}$ for other values of $q$, as we described in Fig.[\ref{plot2}].

\subsection*{Acknowledgments}
This work was supported by MINECO (Spain), project PID2019-104397GB-I00 (SDO). The work of KB was supported in part by the JSPS KAKENHI Grant Number 
JP21K03547. This research was also supported in part by the International Centre for Theoretical Sciences (ICTS) 
for the online program - Physics of the Early Universe: ICTS/peu2022/1 (TP).

\end{document}